\documentclass[]{pasj01}
\newcommand{\kms}{{\rm km\,s^{-1}}}

\usepackage{natbib,color,graphicx}
\begin{document} 
\Received{}
\Accepted{}

\title{The Detection Rates of Merging Binary Black Holes Originating from Star Clusters and Their Mass Function}

\author{Michiko S. \textsc{Fujii}\altaffilmark{1}}%
\altaffiltext{1}{Department of Astronomy, Graduate School of Science, The University of Tokyo, \\7-3-1 Hongo, Bunkyo-ku, Tokyo, 113-0033, Japan}
\email{fujii@astron.s.u-tokyo.ac.jp}

\author{Ataru \textsc{Tanikawa},\altaffilmark{2}}
\altaffiltext{2}{Department of Earth Science and Astronomy, College of Arts and Sciences,
The University of Tokyo, \\3-8-1 Komaba, Meguro-ku, Tokyo 153-8902, Japan}

\author{Junichiro \textsc{Makino}\altaffilmark{3}\altaffilmark{4}}
\altaffiltext{3}{Department of Planetology, Graduate School of Science, Kobe University,\\
1-1, Rokkodai-cho, Nada-ku, Kobe, Hyogo, Japan, 657-8501}
\altaffiltext{4}{RIKEN Advanced Institute for Computational Science,\\ 7-1-26 Minatojima-minami-machi, Chuo-ku, Kobe, Hyogo 650-0047, Japan}



\maketitle

\begin{abstract}
Advanced LIGO achieved the first detection of the gravitational wave, 
which was from a merging binary black hole (BBH). In the near future, 
more merger events will be observed, and the mass distribution of
them will become available. The mass distribution of merger 
events reflects the evolutionary path of BBHs: 
dynamical formation in dense star clusters 
or common envelope evolution from primordial binaries. 
In this paper, we estimate the detection rate of merging BBHs which 
dynamically formed in dense star clusters by combining
the results of $N$-body simulations, modeling of globular clusters, 
and cosmic star-cluster formation history. 
We estimate that the merger rate density in the local universe within 
the redshift of 0.1 is 13--57\,Gpc$^{-3}$\,yr$^{-1}$.
We find that the detection rate is 0.23--4.6 per year for the current
sensitivity limit and that it would
increase to 5.1--99 per year for the designed sensitivity
which will be achieved in 2019.
The distribution of merger rate density in 
the local universe as a function of redshifted chirp mass has
a peak close to the low-mass end.
The chirp mass function of the detected mergers, 
on the other hand, has a peak at the 
high-mass end, but  is almost flat. This difference is simply because
the detection range is larger for 
more massive BBHs.

\end{abstract}


\section{Introduction}

The detection of the gravitational wave (GW) source GW150914 by the
Laser Interferometer Gravitational Wave Observatory (LIGO) was
interpreted as the merger of a binary black hole (BH) whose members
have masses of $m_1 = 36^{+5}_{-4}M_{\odot}$ and $m_2 =
29^{+4}_{-4}M_{\odot}$ \citep{2016PhRvL.116f1102A}. The gravitational wave has
finally become the reality, and advanced LIGO (aLIGO) opens up a new 
window to observe the universe.

What we have seen through this newly opened window is quite
a big surprize in the sense that it was not a merger event of a binary
neutron star (NS). Previous studies of the event rate before the detection
of GW150914 generally concluded that the detection rate of NS-NS
mergers is higher than that of BH-BH mergers by more than one order of
magnitude \citep[e.g.,][]{2010CQGra..27q3001A}.

One obvious reason that researchers did not expected the merging of
two 30-$M_{\odot}$ BHs as the dominant sources of GW events is simply
that there was no observational evidence for BHs with a mass more
than $10 M_{\odot}$, except for those of supermassive BHs at the
centers of galaxies and only a few evidences of intermediate-mass
black holes (IMBHs) \citep{2009Natur.460...73F}.

On the other hand, once found, it looks rather natural 
that a BBH merger was detected rather than NS-NS mergers.
The detected GW event (35--250 Hz) \citep{2016PhRvL.116f1102A} is in
the frequency range for which aLIGO has high sensitivity. The merging of two 
30-$M_{\odot}$ BHs generates GW with 20 times larger amplitude, 
compared to that
generated by the merging of two 1.4-$M_{\odot}$ NSs.
Even if the
sensitivity is the same, the limit distance for the detection of
merging of two 30-$M_{\odot}$ BHs is 20 times larger than that for
NS-NS mergers. When converted to the search volume, the difference is
as large as $10^4$. Therefore, if the event rate of 
30-$M_{\odot}$ BH mergers is larger than $10^{-4}$ of the rate of NS-NS
mergers,  GW events observed by aLIGO and other ground-based
detectors will be dominated by binaries with 30\,$M_{\odot}$ or more 
massive BHs. In fact, the distance of GW150914 is 400\,Mpc, while the
current detection limit of aLIGO for NS-NS mergers is only around 80\,Mpc
\citep{2016PhRvL.116f1102A,2016LRR....19....1A}.  
There is nearly an order of magnitude difference between
the actual distance of the observed GW event and the current limit for
NS-NS merger events.
Previous theoretical studies indeed predicted the detection rate of BBH mergers
is roughly two orders of magnitude higher than that of NS-NS mergers 
\citep{2014MNRAS.440.2714B,2015ApJ...806..263D}.
In addition, recent stellar evolution models with  updated stellar wind models
suggest that massive BHs with a mass of up to $80M_{\odot}$ easily form
from single massive stars
\citep{2008NewAR..52..419V}. Some recent studies have predicted
that massive BBH
mergers would be detected more frequently than NS-NS mergers 
\citep{2010ApJ...714.1217B,2011MNRAS.416..133D,2013MNRAS.429.2298M,
  2015PhRvL.115e1101R}.

What GW150914 tells us are (a) the gravitational wave from
merging of BHs really exists and is observed, and (b) there are
BHs with a mass more than $10 M_{\odot}$, and some of them do form
binaries which merge within the Hubble time. 
From the observational point of view, we can expect that a large number
of event in this mass range will be observed, and the precise mass function
and even the redshift distribution will be determined 
in the near future. 
From the theoretical point of view, key questions posed by
GW150914 are how 30-$M_{\odot}$ BHs formed and how they ended up in
binaries and eventually merged. A related important question is whether
they fill the gap between observed stellar-mass BHs and
supermassive BHs. 

Traditional models for the formation
channels of BBHs are (a) common envelope evolution of primordial 
binary massive stars \citep{2007ApJ...662..504B} and (b) dynamical 
formation in dense star clusters \citep{2000ApJ...528L..17P}. 
Recently, rather exotic
formation models such as the fission of the degenerated core during
the gravitational collapse has been proposed \citep{2016ApJ...819L..21L}, 
but the validity of such models is yet to be confirmed.

For the dynamical formation scenario, \citet{2013MNRAS.435.1358T} 
performed a series of $N$-body simulations of star clusters and
modeled BBH merger history based on the results of the simulations. 
They estimated the merger rate of BBHs originating from globular 
clusters which were born either 10 or 12\,Gyr ago, assuming a number 
density of globular clusters in the universe 
\citep[similar studies have also been done by][]
{2006ApJ...637..937O,2010MNRAS.402..371B}. 
In this paper, we present an improved estimate for the
merger rate, detection rate, mass function, and the redshift dependence
for BBH mergers, using the updated sensitivity of current and future
LIGO \citep{Kissel15,2016LRR....19....1A}, global star 
formation history \citep{2014ARA&A..52..415M}, and the 
initial mass function (IMF) of BHs.

Our main findings are summarized as follows.  The dynamical formation
of BBHs in star clusters prefers the formation of binaries of
the most massive BHs with the mass ratio rather close to unity. Thus,
the total event rate (not considering the sensitivity of the detector)
peaks at the chirp mass of around $10 M_{\odot}$, but the event rate
for 30--50 $M_{\odot}$ events is around 1/10 of that for $10
M_{\odot}$. When we take into account the detector sensitivity, the
event rate becomes almost flat for the range of the chirp mass of
10 to 100 $M_{\odot}$. 
The merger rate in the local universe ($z<0.1$)
is estimated to be 13--57 \,Gpc$^{-3}$\,yr$^{-1}$, which is 
consistent with the value estimated from aLIGO 
\citep[9--240\,Gpc$^{-3}$\,yr$^{-1}$ ][]{2016arXiv160604856T}.
The detection rates are estimated to be 
0.23--4.6 and 5.1--99 per year for the current
and future sensitivity limits, respectively.

\section{Methods}

We estimate the merger rate of BBH dynamically formed 
in star clusters, combining 
BBH merger histories modeled from 
the results of $N$-body simulations of star clusters 
\citep{2013MNRAS.435.1358T}
with a globular cluster formation history following a 
cosmic star formation rate. We follow the method of 
\citet{2013MNRAS.435.1358T} 
\citep[see also][]{2006ApJ...637..937O,2011MNRAS.416..133D}, but
we adopt a time-depending number density of globular 
clusters ($n_{\rm GC}$) assuming that the number of forming star 
clusters is proportional to the cosmic star formation rate. 
We finally calculate the detection rate of BBH merger
events by means of aLIGO assuming
a power spectral density of aLIGO
  \citep{Kissel15,2016LRR....19....1A}. 
In the following, we describe the details of our methods.

\subsection{Model for BH merger history per globular cluster}

\citet{2013MNRAS.435.1358T} performed a series of $N$-body 
simulations of isolated globular clusters and modeled 
the merger event history of the escaping BBHs. In globular 
clusters, the separation of binaries shrink due to three-body
interactions, 
and if they become tight enough, they are ejected 
from the clusters. Once BBHs are ejected, they experience 
no more encounter. The orbital parameters when they escape 
determine their merger timescale due to GW radiation. From 
the escaping rates and the distribution of orbital parameters of 
escaping BBHs, we can estimate the merger rate of BBHs 
dynamically formed in globular clusters.
We here summarize the $N$-body simulations and modeling 
in \citet{2013MNRAS.435.1358T} and then describe our 
modified model.

\subsubsection{Summary of \citet{2013MNRAS.435.1358T}}
\begin{figure}
\begin{center}
  \includegraphics[clip, width=1.\columnwidth]{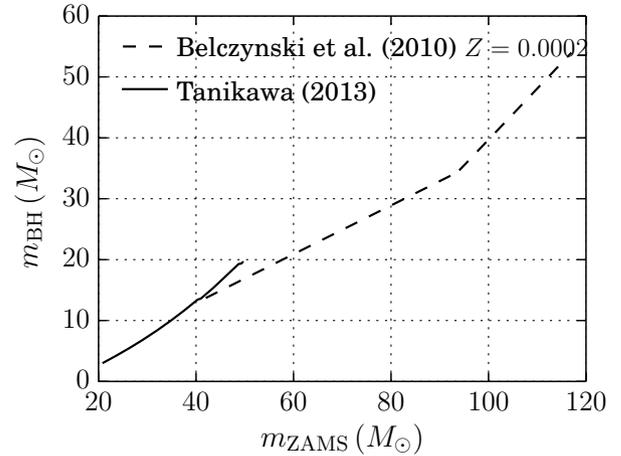}
 \end{center}
\caption{Relation between ZAMS and BH masses adopted in \citet{2013MNRAS.435.1358T} (full line) and obtained by \citet{2010ApJ...714.1217B} (dashed line).\label{fig:mBH}}
\end{figure}

In \citet{2013MNRAS.435.1358T}, a series of 
direct $N$-body simulations 
of star clusters was performed for three different values of the
initial densities of star clusters.
The initial mass functions were given by Kroupa's mass 
function \citep{2001MNRAS.322..231K} with a mass range 
of 0.1--50$M_{\odot}$, and the initial 
distribution of stars followed a King model 
\citep{1966AJ.....71...64K} with a dimensionless concentration
parameter $W_0=7$. Simulations were performed using 
{\tt NBODY4} \citep{2003gnbs.book.....A}, in which stellar and binary 
evolution models \citep{2000MNRAS.315..543H,2001MNRAS.323..630H} 
were included. For massive stars, however, a model 
suggested by \citet{2004MNRAS.353...87E} was used. 
The metallicity was assumed to be $Z=0.001$.
The relation between the zero-age main-sequence (ZAMS) stellar mass
and the resulting BH mass is shown in Figure \ref{fig:mBH}
\citep[same as Figure 1 in ][]{2013MNRAS.435.1358T}.
With these models, the maximum mass of BHs is $\sim 20M_{\odot}$.
In Figure \ref{fig:mBH_m20}, we present the BH mass function resulting
from these models.

The number of particles of the simulations were $N = 8000$--$10^5$
because it is still difficult to perform a large-$N$ simulations 
up to $N\sim 10^6$. In \citet{2013MNRAS.435.1358T}, instead,  
they confirmed that the cumulative number of BBH escaping from 
star clusters is written as a function of the 
thermodynamical time, $\tau$, which is given by
\begin{eqnarray}
\tau = \int _{0}^{t} \frac{dt'}{t_{\rm rh}},\label{eq:tau}
\end{eqnarray}
where $t_{\rm rh}$ is half-mass relaxation time given by
\begin{eqnarray}
t_{\rm rh}=0.0477\frac{N}{\sqrt{G\rho_{\rm h}}\log(0.4N)}
\end{eqnarray}
\citep{1987degc.book.....S}.
Here, $\rho_{\rm h}$ and $N$ are the half-mass density and the 
number of particles, respectively.
\citet{2013MNRAS.435.1358T} found that 
the cumulative number of BBH escapers irrespective of BH mass 
is obtained from the $N$-body simulations as 
\begin{eqnarray}
N_{\rm BBH, esc}(\tau) = 0.020N_{\rm BH,i}\tau,
\end{eqnarray}
where $N_{\rm BH,i}$ is the number of BHs obtained from the IMF.
Using this function, we can estimate a merger rate per cluster
with a given number of particles.

In \citet{2013MNRAS.435.1358T}, three models with different 
numbers of particles ($N=5\times10^5$, $10^6$, and $2\times 10^6$)
 are adopted. They correspond to 
$3.1\times 10^5$, $6.1\times 10^5$, and $1.2\times 10^6M_{\odot}$,
respectively assuming the mean stellar mass is $0.61M_{\odot}$. 
We call these models small, middle, and large, respectively. 
We adopt an initial half-mass density of 
$10^5M_{\odot}{\rm pc}^{-3}$, which is typical for globular clusters 
\citep{2008MNRAS.389.1858H,2009MNRAS.395.1173G,2011MNRAS.410.2698G}
and the middle density model in \citet{2013MNRAS.435.1358T}.
We summarize the models of star clusters in Table \ref{tb:models}.

\begin{table}
  \tbl{Models of star clusters}{%
  \begin{tabular}{lcccc}
\hline
   Model & $N$ & $m_{\rm BH, max} (M_{\odot})$  &  $M_{\rm cl}(M_{\odot})$ & $N_{\rm BH, i}$ \\\hline
   Small ($54M_{\odot}$)& $5\times 10^5$ & 54& $3.1\times 10^5$ & 919\\
   Small ($20M_{\odot}$)& $5\times 10^5$ & 20 & $3.1\times 10^5$ & 715\\
   Middle ($54M_{\odot}$)& $10^6$ & 54 & $6.1\times 10^5$ & 1839 \\
   Middle ($20M_{\odot}$)& $10^6$ & 20 & $6.1\times 10^5$ & 1430\\
   Large ($54M_{\odot}$)& $2\times 10^6$ & 54 & $1.2\times 10^6M_{\odot}$ & 3677\\ 
   Large ($20M_{\odot}$)& $2\times 10^6$ & 20 & $1.2\times 10^6M_{\odot}$ & 2860\\ 
\hline
  \end{tabular}}\label{tb:models}
  \begin{tabnote}
    For all models, $W_0=7$ and $\rho_{\rm h}=10^5M_{\odot}$\,pc$^{-3}$ are used.
  \end{tabnote}
\end{table}

From the results of the $N$-body simulations, 
\citet{2013MNRAS.435.1358T} also obtained  
the primary mass of BBHs ($m_1$), mass ratio ($q=m_2/m_1$), eccentricity, 
and semi-major axis 
distributions of escaping BBHs. They confirmed that these do not depend 
on $N$. They modeled these distributions by fitting a function to 
the results of their simulations.
We hereafter describe their models.

The probability distribution as a function of BH mass ratio ($q$), 
eccentricity ($e$), and semi-major axis ($a$) 
obtained from the simulations are expressed as
\begin{eqnarray}
P_1(q)= \left\{
      \begin{array}{l}
      0 \quad ({\rm for}\; 0<q<0.5),\\     
      \int_{0.5}^{q}2 dq' \quad ({\rm for}\; 0.5<q<1) 
      \end{array}
      \right.
\label{eq:q}
\end{eqnarray}
\citep[see equations (12) and (13) in ][]{2013MNRAS.435.1358T},
\begin{eqnarray}
P_2(e) = \int _{0}^{e}2e'de'
\label{eq:e}
\end{eqnarray}
\citep[see equations (10) and (11) in ][]{2013MNRAS.435.1358T},
and 
\begin{eqnarray}
P_3(a) = \int _{0}^{a}p_{3}(a')da',
\label{eq:a}
\end{eqnarray}
where
\begin{eqnarray}
p_3(a) = \frac{1}{\sqrt{2\pi}\sigma(a/a_{1kT})} 
\exp \left[ -\frac{1}{2\sigma^2}
\left\{ \log \left(\frac{a}{a_{1kT}}\right) -\log\mu \right\}^2 
\right]
\end{eqnarray}
\citep[see equations (8) and (9) in ][]{2013MNRAS.435.1358T}, respectively.
Here, $\sigma=0.81$ and $\mu=0.15$ \citep{2013MNRAS.435.1358T}, 
and $a_{1kT}$ is a semi-major axis
with which the binding energy of a binary is $1kT$, where $3/2kT$ is 
the initial average kinetic energy of stars in the star cluster.
The relation between the primary mass of BBH mass ($m_1$) and 
the escape time of the BBH in thermodynamical time of the 
cluster ($\tau$) is given by
\begin{eqnarray}
m_1({\tau}) = 
      \left\{
      \begin{array}{l}
      20M_{\odot}  \quad (0.5<\tau<1.5),\\
      20M_{\odot}\left( \frac{\tau}{1.5}\right) ^{-1} \quad (\tau>1.5)
      \end{array}
       \right.
\label{eq:m1}
\end{eqnarray}
\citep[see equation (14) in ][]{2013MNRAS.435.1358T}.

Using these functions,
we can generate a distribution of escaping BBHs per cluster
by means of a Monte Carlo technique and then 
calculate their merging timescale due to the GW radiation,
which is given by
\begin{eqnarray}
t_{\rm GW} = \frac{5}{256} \frac{c^5}{G^3} \frac{a^4}{m_{1}^3q(1+q)}g(e),
\end{eqnarray}
where 
\begin{eqnarray}
g(e) = \frac{(1-e^2)^{3.5}}{1+(73/24)e^2+(37/96)e^4}.\label{eq:g}
\end{eqnarray}
Here $c$ is the light speed.

\subsubsection{Generating BBH Merger History per Cluster}

\begin{figure*}
 \begin{center}
  \includegraphics[clip, width=0.9\columnwidth]{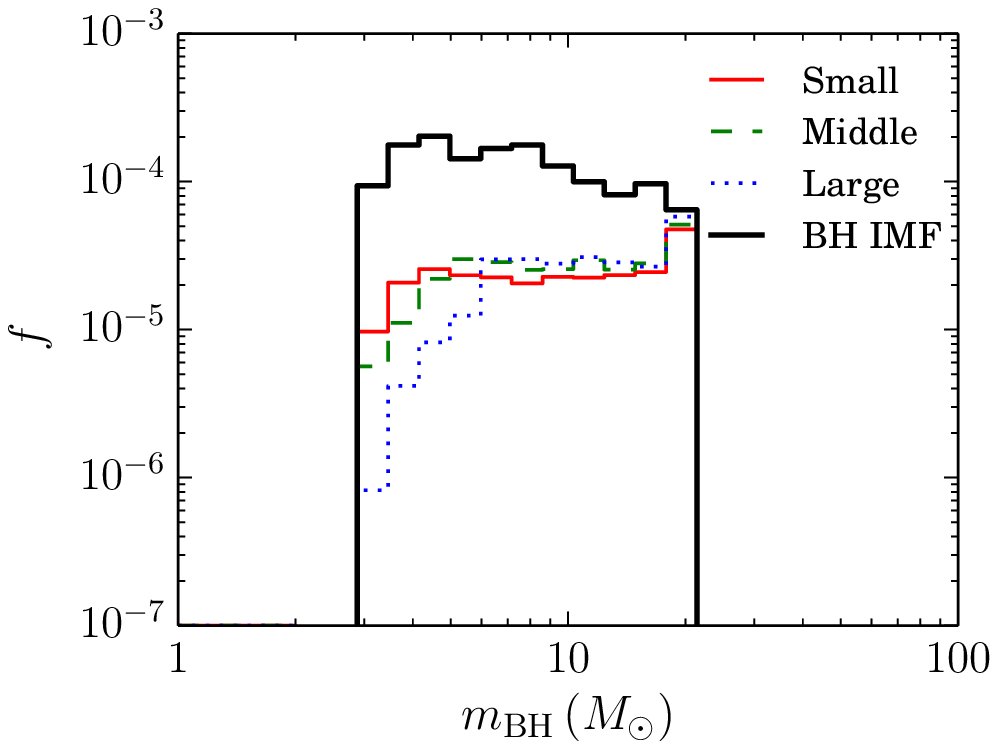}
  \includegraphics[clip, width=0.9\columnwidth]{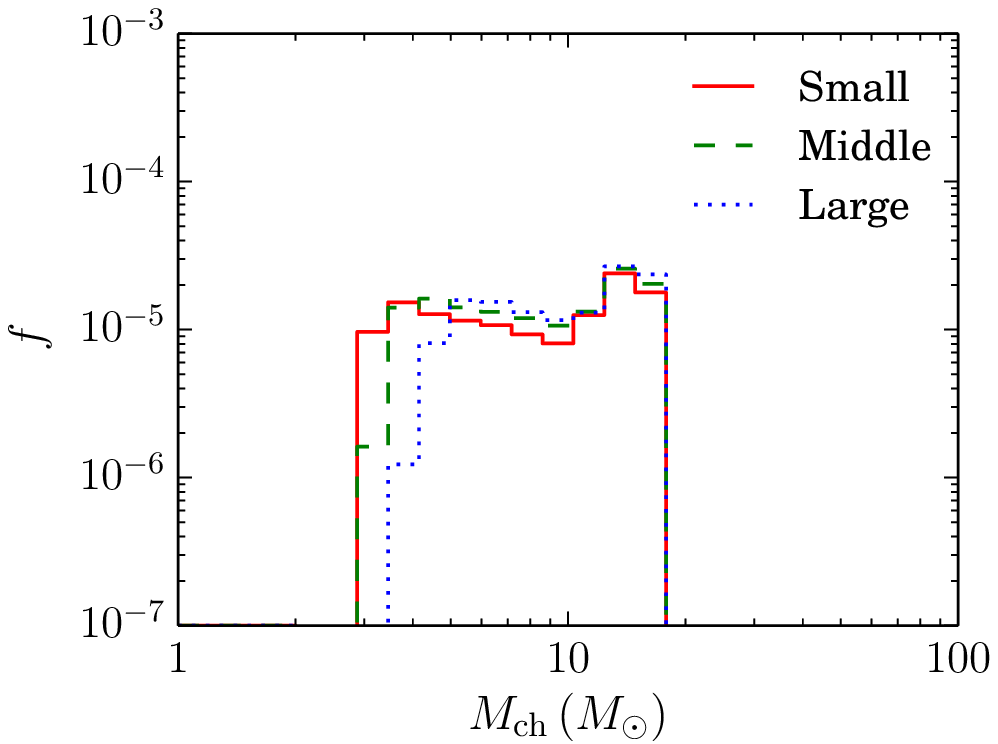}\\
  \includegraphics[clip, width=0.9\columnwidth]{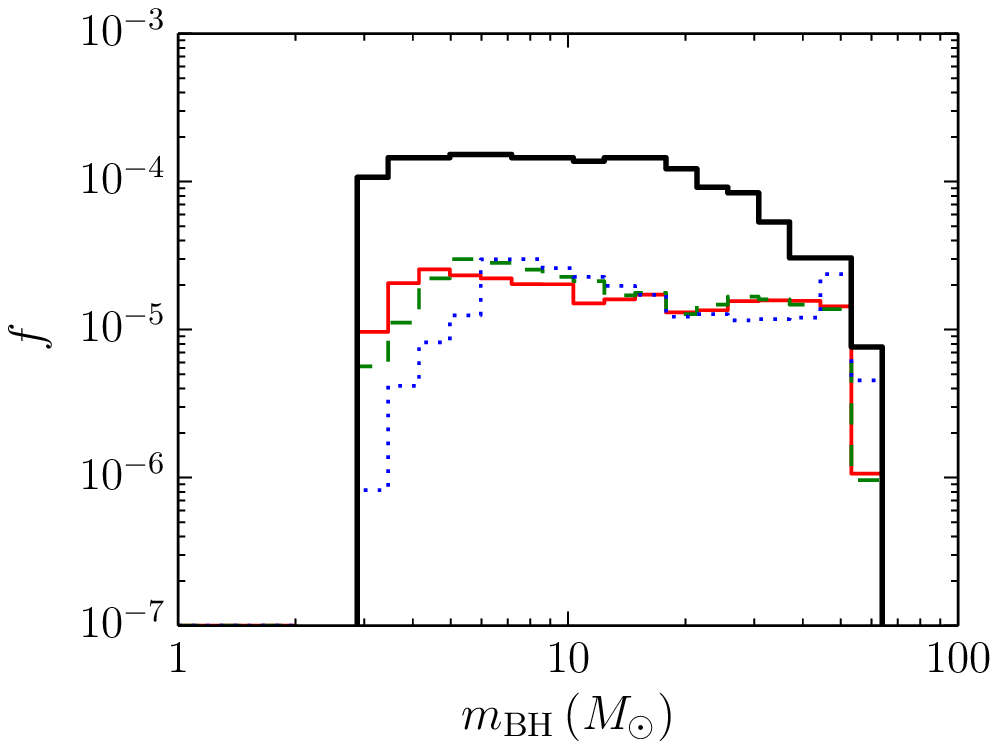}
  \includegraphics[clip, width=0.9\columnwidth]{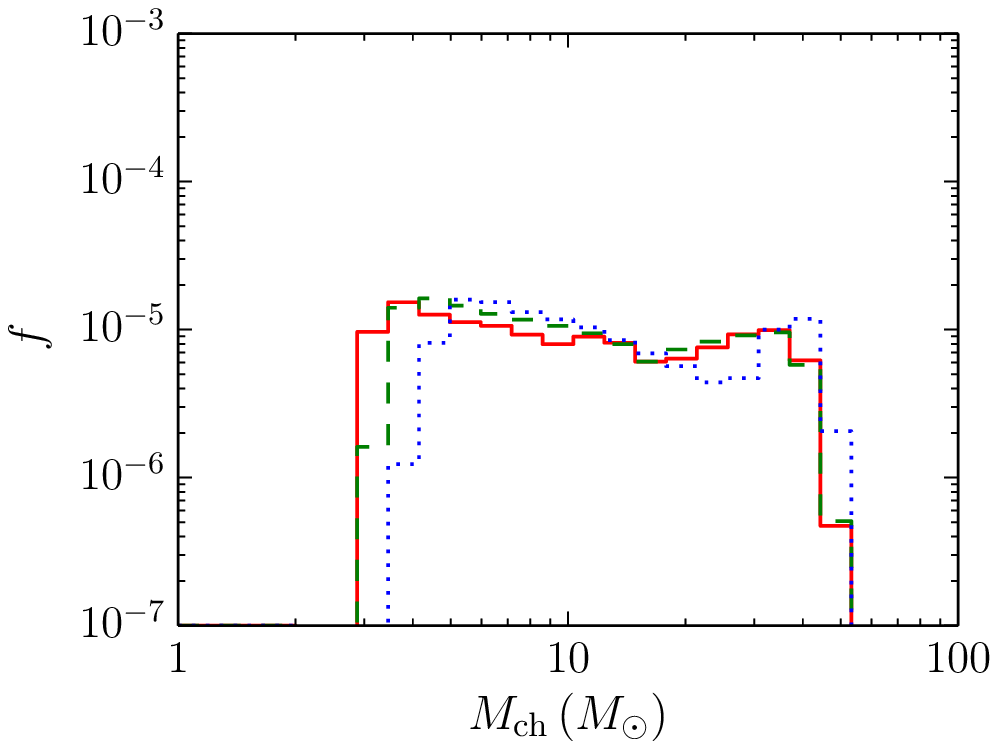}\\
 \end{center}
\caption{Mass function of BHs which contribute to mergers
for models small ($N=5\times10^5$), middle ($N=10^6$), and large 
($N=2\times10^6$) of \citet{2013MNRAS.435.1358T} (top) and us (bottom).
The fraction is normalized by the total number of cluster
particles.  Black histogram show the mass distribution of BHs obtained from the IMF and stellar evolution model.  
Right: normalized chirp mass distribution of merging BBHs for the models
of \citet{2013MNRAS.435.1358T} (top) and us (bottom). \label{fig:mBH_m20}}
\end{figure*}

\begin{figure}
 \begin{center}
  \includegraphics[clip, width=1.\columnwidth]{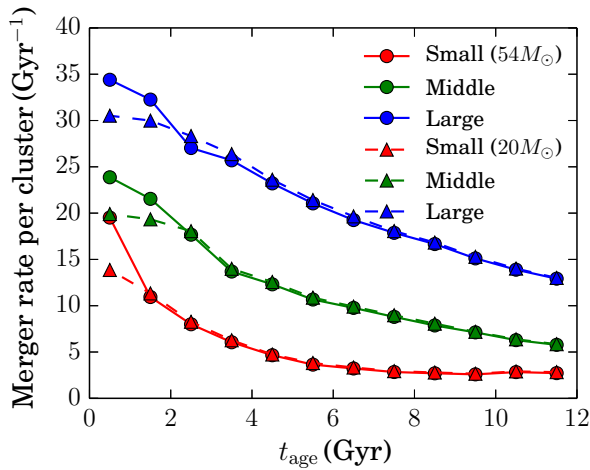}
 \end{center}
\caption{Merger rate per cluster as a function of cluster age 
for models with a maximum mass for 
black holes of $54M_{\odot}$ (dots and full curves) and $20M_{\odot}$
(triangles and dashed curves). \label{fig:merger_rate}}
\end{figure}

Using equations (\ref{eq:tau})--(\ref{eq:g}), 
we construct merger event histories for each cluster model.
In the top left panel of Figure \ref{fig:mBH_m20}, we present the 
mass function of BHs which contribute to BBH mergers 
($m_1$ and $m_2$) up to a cluster age of 12\,Gyr for our three cluster models.
From the stellar evolution models that we adopt, the lower-mass 
limit of BHs is set to be $3M_{\odot}$.
For comparison, we also show the 
BH mass distribution obtained from the adopted IMF and stellar 
evolution model. We confirmed that the numbers of BHs which 
contribute to BBH mergers generated from our model do not exceed 
the number of BBHs expected from the IMF. 
We also present the chirp mass function in the top right panel of 
Figure \ref{fig:mBH_m20}.
The merger rates per cluster per 1\,Gyr are presented
in Figure \ref{fig:merger_rate}. For all models, 
the merger rates decrease with time. 
Although \citet{2013MNRAS.435.1358T} varied the initial density
of globular clusters, we adopt that for their standard model
($10^5M_{\odot}{\rm pc}^{-3}$).

In \citet{2013MNRAS.435.1358T}, the maximum mass of BHs was limited
to $20M_{\odot}$ because their IMF was limited to $50M_{\odot}$ 
(see Figure \ref{fig:mBH}).
On the other hand, the BH masses observed by GW were $\sim 30M_{\odot}$ 
\citep{2016PhRvL.116f1102A}, and as a consequence, we cannot 
estimate the merger rates of such massive BBHs with their model. 
We therefore modify the model given by \citet{2013MNRAS.435.1358T}
to include more massive BHs. Equation (\ref{eq:m1}) gives  
the relation between the primary mass of BBH mass ($m_1$) and 
the escape time of the BBH ($\tau$). This relation shows that 
in star clusters with a realistic mass function, massive stars sink to 
the cluster core quickly and tend to form hard binaries due to the mass 
segregation. As a result, more massive binaries escape earlier from 
the star cluster. 
Since $N$-body simulations performed by \citet{2013MNRAS.435.1358T}
had an upper limit of the BH mass of $20M_{\odot}$, all escapers 
in $\tau<1.5$ is assumed to have the maximum BH mass. We change 
this assumption to that the escape time of the BBH ($\tau$) 
continuously decreases up to a new maximum BH mass because 
if star clusters contain more massive BHs, it is natural to 
consider that they have shorter escape time. 
We therefore extrapolate equation (\ref{eq:m1}) to a new upper-mass 
limit ($54M_{\odot}$), 
which is the maximum BH mass which can avoid the pair instability
supernova range \citep{2014ApJ...789..120B}. This new relation 
is expressed as
\begin{eqnarray}
m_1({\tau}) = 
      \left\{
      \begin{array}{l}
      54M_{\odot}  \quad (0.5<\tau<0.55),\\
      20M_{\odot}\left( \frac{\tau}{1.5}\right) ^{-1} \quad (\tau>0.55).
      \end{array}
       \right.
\label{eq:m1_54}
\end{eqnarray}
The distributions of the generated merging BH mass
and the chirp mass are presented in the bottom panels of 
Figure \ref{fig:mBH_m20}.
We assume that the probability distributions for $q$, $e$, and $a$
do not change. 
The obtained merger rates are shown in Figure 
\ref{fig:merger_rate}. At 0--1\,Gyr, the merger rates are slightly 
higher than those for the models 
with a maximum BH mass of $20M_{\odot}$. This is because  
more massive BHs have a shorter merger time. 

In order to compare these results with models that directly 
constructed from the results of $N$-body simulations, we also present 
the BH mass function obtained from IMF up to 120$M_{\odot}$ in the 
bottom left panel of Figure \ref{fig:mBH_m20}. Here, 
we assume a stellar evolution model which gives a relation between
ZAMS and BH masses shown in Figure \ref{fig:mBH} \citep{2010ApJ...714.1217B}.
Here, $Z=0.0002$ is assumed. 
For $m_{\rm ZAMS}<42M_{\odot}$, the result is consistent 
with that of \citet{2013MNRAS.435.1358T}.
For simplicity, we adopted 
\begin{eqnarray}
m_{\rm BH} (M_{\odot})= 
      \left\{
      \begin{array}{l}
      0.52(M_{\rm ZAMS}/1M_{\odot}) - 8.1 \quad (21<M_{\rm ZAMS}/1M_{\odot}<42)\\
      0.40(M_{\rm ZAMS}/1M_{\odot}) - 3.1 \quad (42<M_{\rm ZAMS}/1M_{\odot}<94)\\
      0.84(M_{\rm ZAMS}/1M_{\odot}) - 44  \quad (94<M_{\rm ZAMS}/1M_{\odot}<120)
      \end{array}
       \right .
\label{eq:mBH_mZAMS}
\end{eqnarray}
which is obtained by a least-squares fitting to the models of 
\citet{2010ApJ...714.1217B} and \citet{2013MNRAS.435.1358T}. 
The small discrepancy around 40--50$M_{\odot}$ in the ZAMS mass 
comes from the difference in the assumed stellar evolution models. 
In the bottom left panel of Figure \ref{fig:mBH_m20}, 
we confirm that the numbers of BHs which contribute to BBH mergers
do not exceed that expected from IMF.

In the following,
we discuss the relation between the BH mass function (MF) obtained
from IMF and the mass distribution of BHs contributed to merger
events. The relation between the thermodynamical time (similar to merger
timescale) and primary BH mass obtained from the results of $N$-body
simulations is $\tau \propto m_1$. 
On the other hand, the BH MF 
obtained from IMF (see the left panels of Figure \ref{fig:mBH_m20}) 
follows $dN/dm_{\rm BH}\propto m_{\rm BH}^{-1}$
for the higher mass region. Combining these, we expect that 
the MF of BH mergers becomes almost flat at 
the high mass end. Indeed, we see almost flat distribution.
The flat region becomes wider as the relaxation time decreases.
In our models, model Small has the shortest relaxation time, 
and therefore it dynamically evolves most quickly.
During the dynamical evolution of star clusters, the most massive 
BBHs selectively form and are ejected in the beginning,
and then less massive BBHs start to be ejected. Therefore, the
fraction of low-mass BBH 
mergers increases for clusters with a smaller $N$.
Thus, our model reasonably reflects the dynamical evolution of 
star clusters and the distribution of BBHs formed there.
We note that the cut-off mass at $m_{\rm BH}=3M_{\odot}$ is adopted
only in this paper, and therefore the BH mass distributions 
of this paper and \citet{2013MNRAS.435.1358T} are different
at the low-mass end.

Natal kicks due to assymmetric supernova explosions also affect
the merger rates of BBHs in star clusters because some BHs are 
ejected from their host star clusters due to the high-velocity
kicks. The retention fraction is estimated to be 0.1 
for neutron stars \citep{2002ApJ...573..283P}, but that of BHs
depends on the fraction of fallback materials when they collapse,
which affects the kick velocity 
\citep{2012ApJ...749...91F, 2016arXiv160202444R}.
The fallback fraction increases as the BH mass increases,
and as a result the kick velocity decreases 
\citep{2012ApJ...749...91F}. All BHs with a progenitor mass
of $>40M_{\odot}$ are expected to retain in the cluster 
\citep{2016arXiv160202444R}. 
In \citet{2013MNRAS.435.1358T}, the effect of natal kicks are 
included in their $N$-body simulations by a single retention 
fraction irrespective of stellar masses, 
and they found that the BBH merger rates are simply proportional 
to the retention fraction. From this result, the natal kicks
do not seem to affect the dynamical evolution of star clusters.
Therefore, 
we first assume that all BHs retain in star clusters and discuss
the effect of natal kicks and retention rate in Section 4.1.

\subsection{Cosmic star-cluster formation history}

In order to estimate the BBH merger rate for the entire universe,
we need the number density of globular clusters in the universe.
In \citet{2013MNRAS.435.1358T}, a single formation epoch for all
globular cluster was assumed,
but we adopt a number density of globular 
clusters as a function of redshift, $z$, (or universe age, $t$).
We assume that the star-cluster formation density 
is proportional to the star formation density of 
the universe (cosmic star formation history).
Combining all survey data from ultraviolet to infrared, 
\citet{2014ARA&A..52..415M} 
obtained a cosmic star formation history as follows:
\begin{eqnarray}
\psi (z) = 0.015 \frac{(1+z)^{2.7}}{1+[(1+z)/2.9]^{5.6}}
\, {\rm M_{\odot}\, year ^{-1}\, Mpc^{-3}}.
\end{eqnarray}

For the estimation of the number density of 
forming star clusters, we use
the number of star clusters per stellar mass ($10^9M_{\odot}$), $T$,
\citep{1993MNRAS.264..611Z} observationally determined for galaxies. 
We adopt $T$ from the results 
of the Spitzer Survey of Stellar Structure in Galaxies (S$^4$G)
\citep{2010PASP..122.1397S}, which includes both early- and late-type
samples. According to their results, $T$ depends on galaxy stellar 
mass ($M_{\star}$): 
\begin{eqnarray}
T = \left\{
    \begin{array}{l}
    10^{5.7}(M_{\star}/M_{\odot})^{-0.56}\quad ({\rm for}\;10^{8.5}<M_{\star}<10^{10}M_{\odot})\\
    8.3 \quad ({\rm for}\;10^{10}<M_{\star}<10^{11}M_{\odot})\\
    10^{-6.11}(M_{\star}/M_{\odot})^{0.63}\quad ({\rm for}\;M_{\star}>10^{11}M_{\odot})
    \end{array}   
    \right.
\end{eqnarray}
\citep{2015ApJ...799..159Z,2016ApJ...818...99Z}.

Using these relation, the maximum and minimum number of star clusters 
per stellar mass of $10^9M_{\odot}$ are $87$ for $M_{\star}=10^{8.5}M_{\odot}$
and 8.3 for $M_{\star}=10^{10.5}$--$ 10^{11} M_{\odot}$.
In any cases, $T\sim$10--100. 
We hereafter ignore the observed dependence of $T$ on the mass of the 
host galaxy and adopt a fixed values of $T=10$ as
our standard value.

\begin{figure}
 \begin{center}
  \includegraphics[clip, width=0.9\columnwidth]{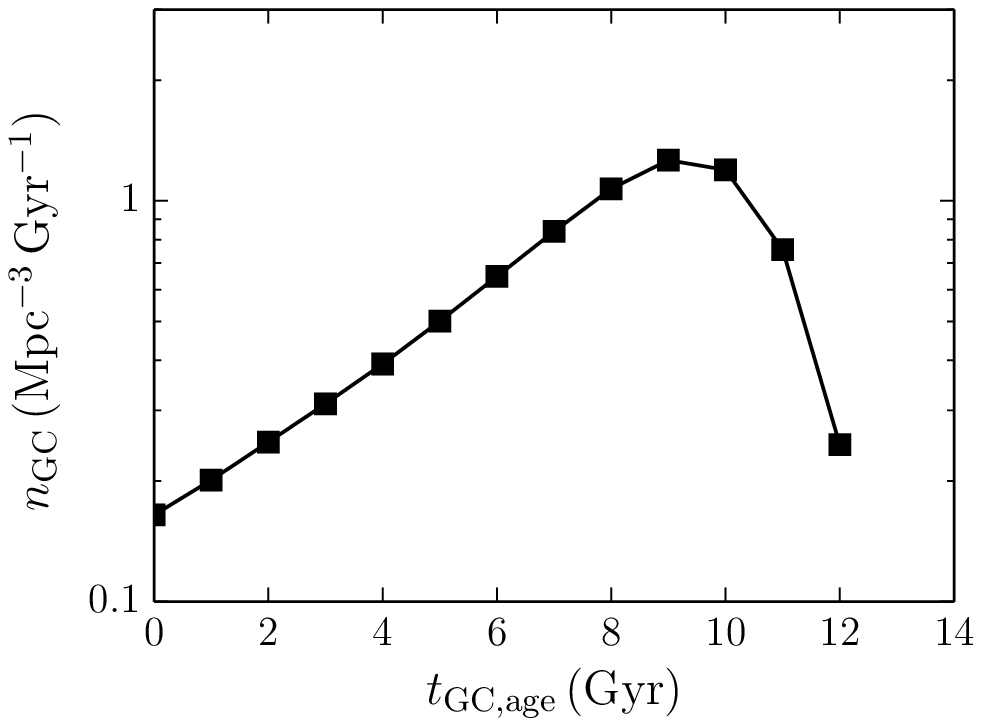}
  \includegraphics[clip, width=0.9\columnwidth]{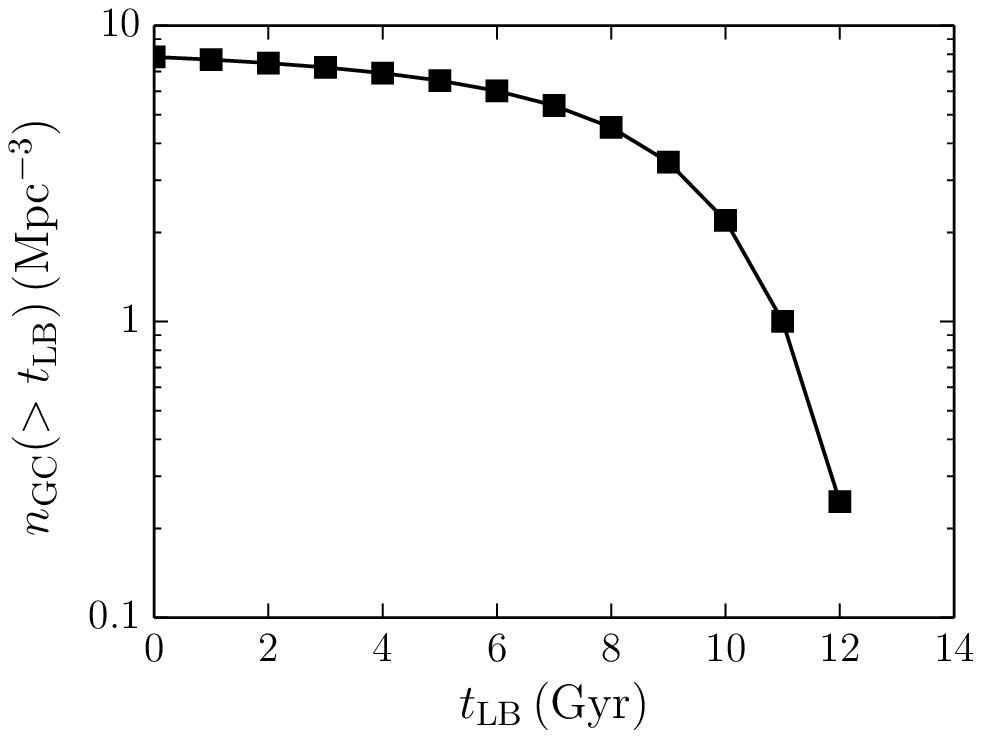}
 \end{center}
\caption{The number density of star clusters in comoving volume as a function of cluster age (top) and the number density of globular clusters born before the lookback time (bottom). Here we adopt $T=10$.}\label{fig:Ngc}
\end{figure}

In the top panel of figure \ref{fig:Ngc}, 
we show the number density of star
clusters in comoving volume as a function of current age of star
clusters ($t_{\rm GC, age}$).
The distribution has a peak at $t_{\rm GC, age}\sim 9$\,Gyr, and the 
number density drops for $t_{\rm GC, age}\gtrsim10$ Gyr. We therefore
ignore star clusters which are born more than 12\,Gyr ago.
In the bottom panel of figure \ref{fig:Ngc}, we show the 
total number density of clusters born before a certain look back
time. The total number density of star clusters is 
$7.8\,{\rm Mpc^{-3}}$ for $T=10$. 
This value is a few times to an order of magnitude larger 
than those estimated in previous studies such as 
$8.4h^3{\rm Mpc^{-3}}$, where $h=H_{0}/100\,\kms$ 
\citep{2000ApJ...528L..17P}, and $0.77 \,{\rm Mpc^{-3}}$
\citep{2015PhRvL.115e1101R}. If we consider star clusters with the
ages 
similar to those of globular clusters (clusters born before 10\,Gyr),
their number density  is $2.2\,{\rm Mpc^{-3}}$
(see Figure \ref{fig:Ngc}),
which is similar to the optimistic estimate in 
\citet{2015PhRvL.115e1101R}.
Although most of massive clusters are globular clusters and 
they are old, in starburst galaxies and star-forming dwarf 
galaxies such as the Large and Small Magellanic Clouds, massive
and dense star clusters, so-called super star clusters, are 
still forming \citep{2010ARA&A..48..431P}.
They are dense and massive enough to form BBHs.
We therefore estimate BBH merger rates both including
and excluding star clusters younger than typical globular
clusters in this study. We discuss the contribution of young 
star clusters to our final results in the following sections.

\subsection{Initial mass function of star clusters}
The mass function of merging BBHs and the merger rate
depend on the total mass of star clusters ($M_{\rm cl}$). We therefore assume an 
initial mass function of globular clusters as 
$dN_{\rm cl}/dM_{\rm cl}\propto M_{\rm cl}^{-2}$ \citep{2010ARA&A..48..431P},
which is determined by the observations of young ($<1$ Gyr) star clusters 
in nearby galaxies. Although this relation is for young
massive clusters in nearby galaxies, we assume that it 
is also applicable to globular clusters. 
Observed current globular cluster mass functions follows a lognormal
distribution with a peak at $\sim 2\times 10^5M_{\odot}$ 
\citep{2006ARA&A..44..193B,2007ApJS..171..101J} 
rather than a power-law.
This lognormal distribution is considered to be shaped from a 
power-low distribution due to the disruption of less massive 
clusters \citep{2001ApJ...561..751F}. 
On the other hand, \citet{2000MNRAS.318..841V} claimed 
that an initially lognormal distribution can reproduce the 
observations better than the power-law model. 
In both cases, star clusters less massive 
than the peak mass do not much contribute the total mass of star 
clusters. 
We therefore take into account only star clusters
with $\gtrsim 2\times 10^5M_{\odot}$.
We set the maximum mass of star clusters to be $\sim 2\times 10^6M_{\odot}$.
If the cluster mass function is lognormal, the mass of the 
most massive cluster is a few times $10^6M_{\odot}$ \citep{2007ApJS..171..101J}.
The power of the observed young clusters is truncated at 
the high-mass end \citep{2010ARA&A..48..431P}. 
We adopt number fractions of our three cluster 
models as 0.57, 0.29, and 0.14 for models small, middle, and large, 
respectively. These fractions follows the power of $-2$. 
We assume that every 1\,Gyr star clusters are born following the cosmic 
star-cluster formation history (see Figure \ref{fig:Ngc}) the number 
fractions.

\subsection{BBH merger rates}

Following \citet{2013MNRAS.435.1358T}, we generate the distributions of 
$100\,000$ BBHs per cluster model using a Monte Carlo technique
\citep[see section 4.3 of][for the details]{2013MNRAS.435.1358T}.
Assuming $n_{\rm GC}(z)$ as described in the previous 
sections, we calculate BBH merger rates originating from 
star clusters in the universe as follows.

The cumulative merger rates up to a given $z$ is 
calculated by integrating 
the merger rate for individual globular clusters 
at a given universe age ($\Gamma _{\rm mrg} (t)$) as:
\begin{eqnarray}
\Gamma _{\rm mrg}(<z) = \int_{0}^{z}\left[ n_{\rm GC}(z')\frac{dV(z')}{dz'}
\frac{\Gamma _{\rm mrg}(z')}{1+z'}\right]dz',
\label{eq:merger_rate}
\end{eqnarray}
where $z$ is a function of $t$ and $1/[1+z(t)]$ is a factor
coming from the cosmological time dilation of the merger rate.
The volume, $dV(z)=dV(t)$, is expressed as:
\begin{eqnarray}
dV(t) = 4\pi D_{\rm p}(t)^2dD_{\rm p}(t),
\end{eqnarray}
where $D_{\rm p}(t)$ is a proper distance, which is obtained by
\begin{eqnarray}
D_{\rm p}(t)=c\int_{t}^{t_0}[1+z(t')]dt',
\end{eqnarray}
where $c$ is the speed of light.
The universe age, $t$, is written as 
\begin{eqnarray} 
t &=& H_{0}^{-1} \int _{z}^{\infty} \frac{dz'}{(1+z')\sqrt{\Omega _{\rm m} (1+z')^3+\Omega _{\Lambda}}}\\
  &=& \frac{1}{3H_{0}\sqrt{\Omega_{\Lambda}}}\log \left[ \frac{\sqrt{\Omega_{\rm m}(1+z)^3+\Omega_{\Lambda}}+\sqrt{\Omega_{\Lambda}}}{\sqrt{\Omega_{\rm m}(1+z)^3+\Omega_{\Lambda}}-\sqrt{\Omega_{\Lambda}}}\right]
\end{eqnarray}
We adopt the $\Lambda$CDM model with 
$H_{0}=67.8{\rm km\,s^{-1}\,Mpc{-1}}$, $\Omega_{\rm m}=0.308$, 
$\Omega_{\rm \Lambda}=0.692$, and $\Omega_{\rm k}=0.0$ from Planck 2015
results \citep{2015arXiv150201589P}.

\begin{figure*}
 \begin{center}
  \includegraphics[clip, width=0.9\columnwidth]{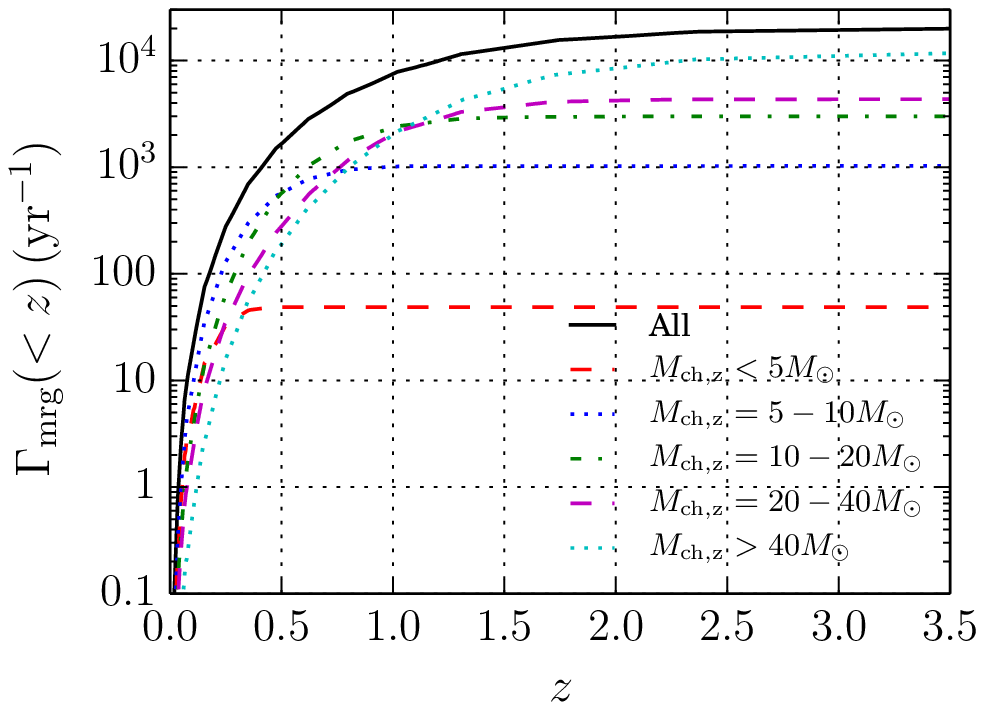}
  \includegraphics[clip, width=0.9\columnwidth]{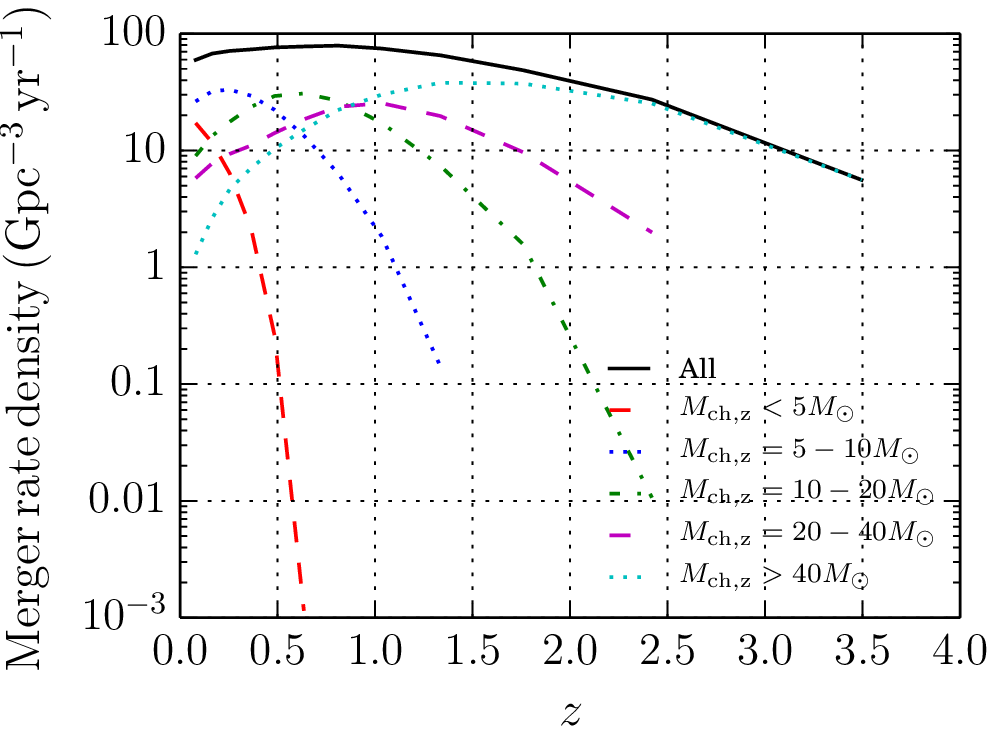}\\
  \includegraphics[clip, width=0.9\columnwidth]{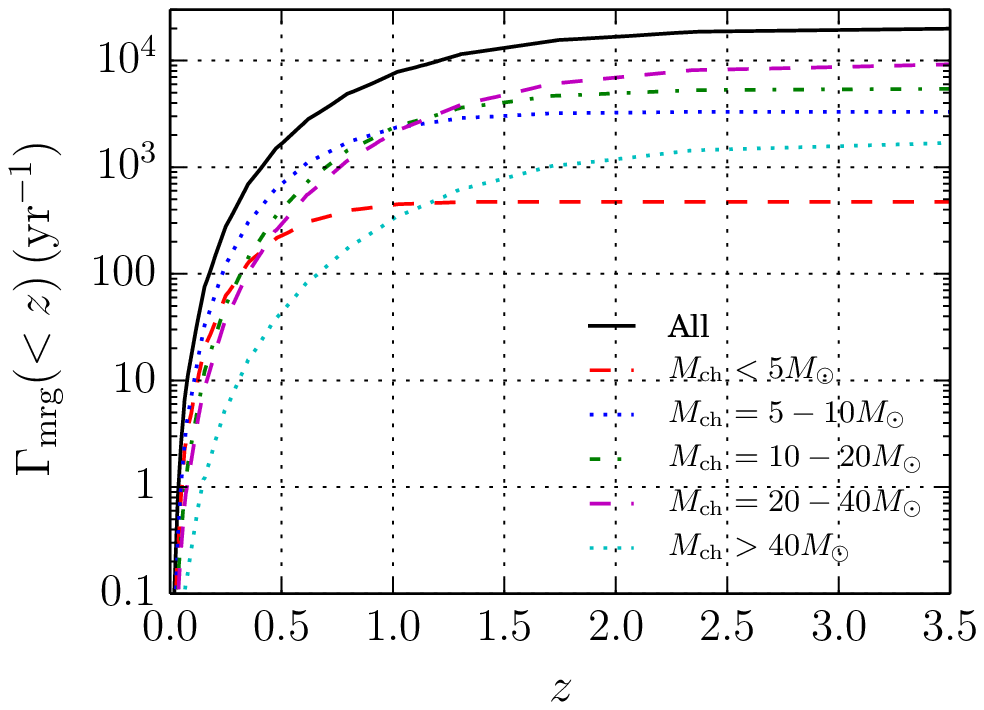}
  \includegraphics[clip, width=0.9\columnwidth]{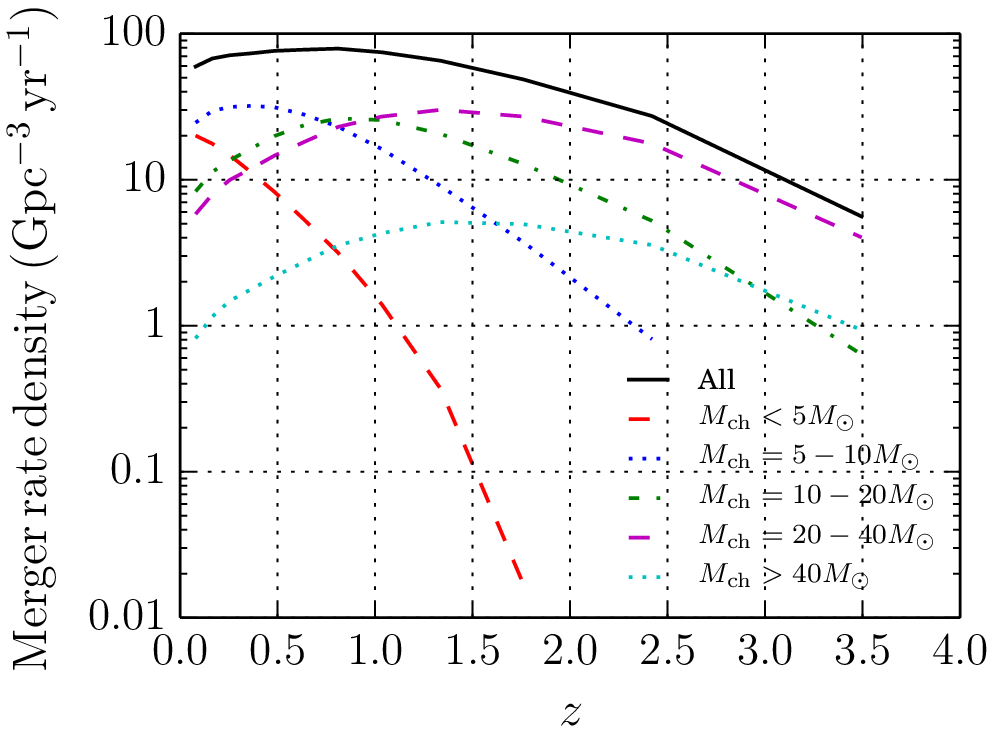}\\
 \end{center}
\caption{Cumulative merger rates (left) and merger rate densities (right). Top and bottom panels are with subsets of redshifted chirp mass and chirp mass, respectively. \label{fig:rate_perV}}
\end{figure*}

We obtain $\Gamma _{\rm mrg}(t)$ from the models based on the $N$-body 
simulations described in Section 2.1 and $n_{\rm GC}(t)$ from the 
cosmic star formation history as described in Section 2.2.
We show the cumulative merger event rate 
obtained from equation (\ref{eq:merger_rate}) and the sub-fractions 
depending on chirp mass and redshifted chirp mass in the left panels
of Figure \ref{fig:rate_perV}.
The merger rates reach to more than $10^4$\,yr$^{-1}$, but not 
all of them are observable. Most of star clusters 
were born 9--12\,Gyr ago and their dynamical activity decreases
with time. Therefore, the majority of BBH mergers occur at high-$z$,
for which LIGO sensitivity is not sufficient.

We also present the merger rate density in the local universe ($z<0.1$)
in the right panels of Figure \ref{fig:rate_perV}.
In the local universe, the BBH merger rate density is estimated to be 
57\,Gpc$^{-1}$ yr$^{-1}$, which is consistent with the value estimated
from LIGO observation, 9--240\,Gpc$^{-3}$\,yr$^{-1}$ 
\citep{2016arXiv160604856T}. If we allow the formation of star clusters
which can form merging BBHs
only in the earlier universe ($>10$\,Gyr ago), the fraction of 
massive BBHs significantly
decreases because massive BBH mergers in the local universe  
originates from younger star clusters in our models. The merger 
rate density decreases down to 13\,Gpc$^{-1}$ yr$^{-1}$ if we assume 
that only old clusters which were born 10--12\,Gyr ago can be the
source of BBH mergers. However, the 
merger rate is still within the observational estimates.

\citet{2016arXiv160202444R} estimated merger rates originating
from BBHs formed in star clusters using the results of Monte-Carlo 
simulations of globular clusters and obtained a merger rate of 
$\sim 5$\,Gpc$^{-3}$\,yr$^{-1}$ from their standard model. 
\citet{2017MNRAS.469.4665P} also estimated a similar merger rate 
density (6.5\,Gpc$^{-3}$\,yr$^{-1}$) using direct $N$-body simulations 
of star clusters
with $5\times 10^4$ particles. 
These values are an order of magnitude lower than ours. 
One possible reason is that while we take all star clusters to $z=0$
into account, they assumed that  only globular clusters form BBHs. 
Therefore, the
total number density of star clusters in our model is an order 
of magnitude higher than their model (see Section 2.2 for the details).

\begin{figure*}
 \begin{center}
  \includegraphics[clip, width=0.9\columnwidth]{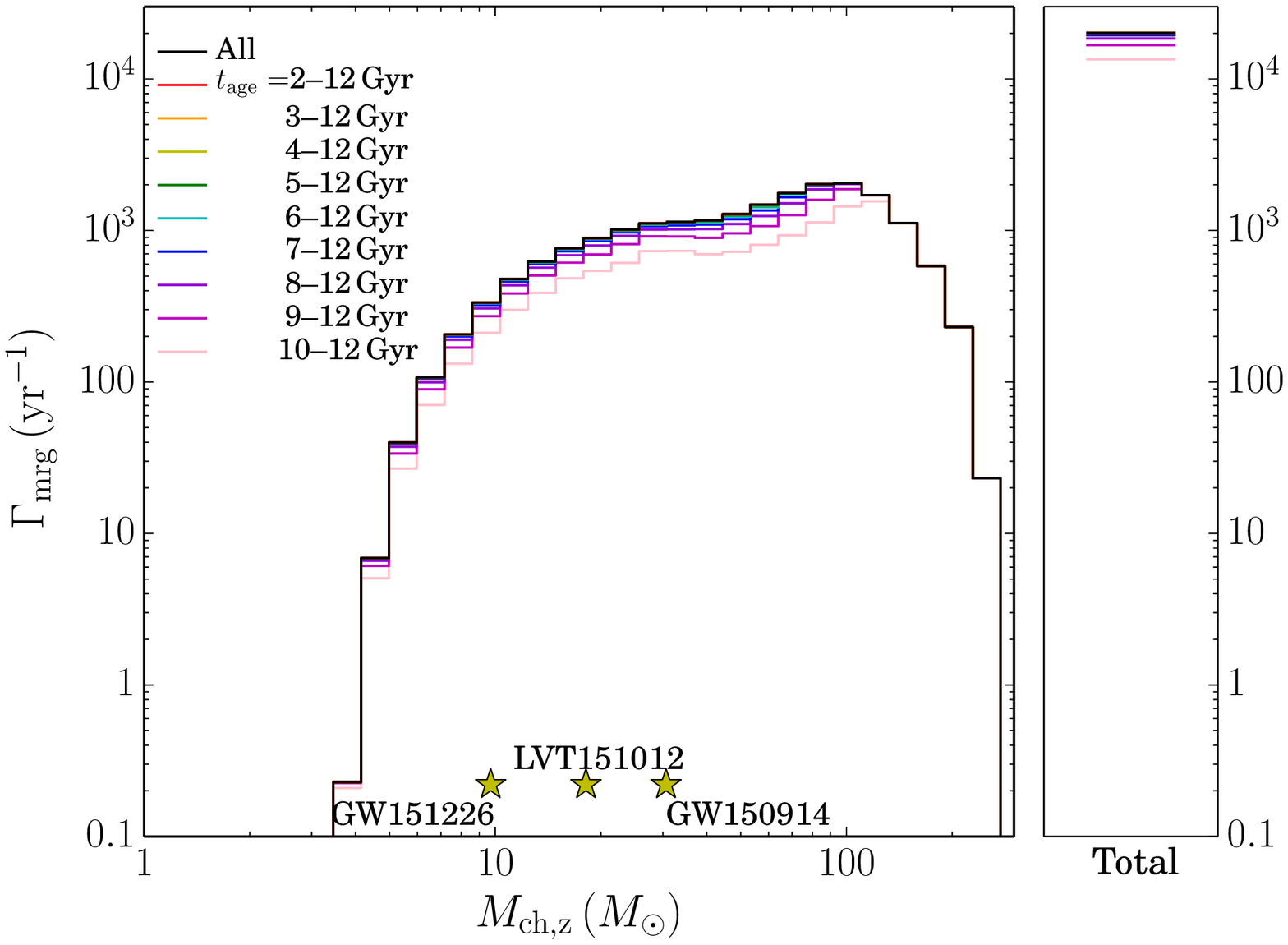}
  \includegraphics[clip, width=0.9\columnwidth]{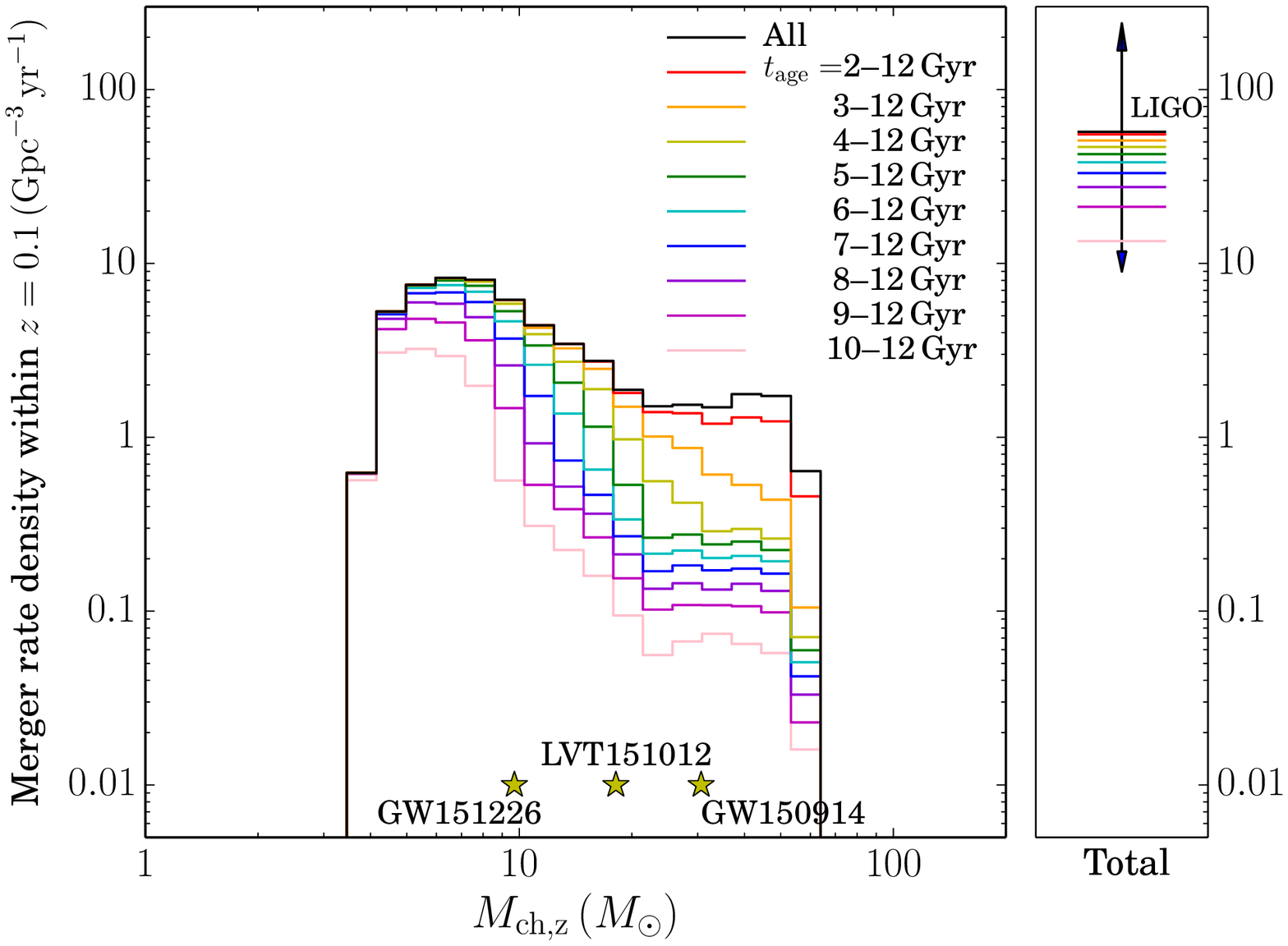}
 \end{center}
\caption{The merger rate distribution as a function of redshifted chirp mass and the total merger rates for all BBH mergers (left) and merger rate density distribution as a function of redshifted chirp mass and the total merger rate density for mergers within $z=0.1$ (right). Black arrow indicate in the right panel indicate the range of the merger rate density estimated by LIGO observation \citep{2016arXiv160604856T}.\label{fig:merger_mass_func}}
\end{figure*}

In the left panel of Figure \ref{fig:merger_mass_func}, 
we present the redshifted chirp mass distribution of BBH mergers
originating from star clusters.  
Including all star clusters, we obtain a weak double-peak 
distribution of redshifted chirp mass. If we exclude young clusters, 
the merger rates slightly decrease, but do not change much because 
most of clusters were born 10--12 Gyr ago.
In the right panel of Figure \ref{fig:merger_mass_func},
we present the merger rate density in the local universe ($z<0.1$)
for comparison with the results of common envelope evolution model
\citep{2016Natur.534..512B}. We here see double-peak distributions.
The high-mass peak is located at $M_{\rm chirp, z}\sim$40--50$M_{\odot}$. 
On the other hand, the peak is the total redshifted mass of 
$\sim $30--40$M_{\odot}$ in the case of common 
envelope model \citep[see Figure 3 in][]{2016Natur.534..512B}.
If we assume equal-mass binaries, it corresponds to 
$M_{\rm chirp, z}\sim 20M_{\odot}$. 
The low-mass peak in the redshifted chirp mass function is at 
5--7\,$M_{\odot}$. The high-mass peak becomes weaker, if we limit the 
cluster formation epoch to higher-$z$.

\subsection{Detectability}

In order to calculate the detection rate, we have to consider 
the detectability of each merger event.
We follow \citet{2013MNRAS.435.1358T} 
\citep[see also][]{2006ApJ...637..937O} and assume that BBH mergers 
are detected when the merger events satisfy:
\begin{eqnarray}
C_{\rm det} = \left( \frac{D_{\rm L}}{D_{\rm L,0}} \right)^{-1} 
\left( \frac{M_{\rm ch, z}}{M_{\rm ch,0}} \right)^{5/6}
\left[ \frac{s(f_{\rm off})}{s(f_{\rm off, 0})} \right]^{1/2} > 1,
\end{eqnarray}
where $D_{\rm L}(=[1+z(t)]D_{\rm p}(t))$ is a luminosity distance,
$M_{\rm ch, z}$ is a redshifted chirp mass, 
and $s(f_{\rm off})$ is a detector response function. 
The redshifted chirp mass is written as 
\begin{eqnarray}
M_{\rm ch,z} = [1+z(t)]M_{\rm ch},
\end{eqnarray}
where $M_{\rm ch}$ is a chirp mass of a binary with masses of 
$m_1$ and $m_2$, which is given by
\begin{eqnarray}
M_{\rm ch} = \frac{(m_1 m_2)^{3/5}}{(m_1 + m_2)^{1/5}}.
\end{eqnarray}
The detector response function is approximated as
\begin{eqnarray}
s(f_{\rm off}) = \int _{0} ^{f_{\rm off}} \frac{(f')^{-7/3}}{S_{\rm N}(f')}df',
\end{eqnarray}
where $S_{\rm N}(f)$ is the noise spectral density 
and $f_{\rm off}$ is the cutoff frequency \citep{1994PhRvD..49.2658C}.

\begin{figure}
 \begin{center}
  \includegraphics[clip, width=1.\columnwidth]{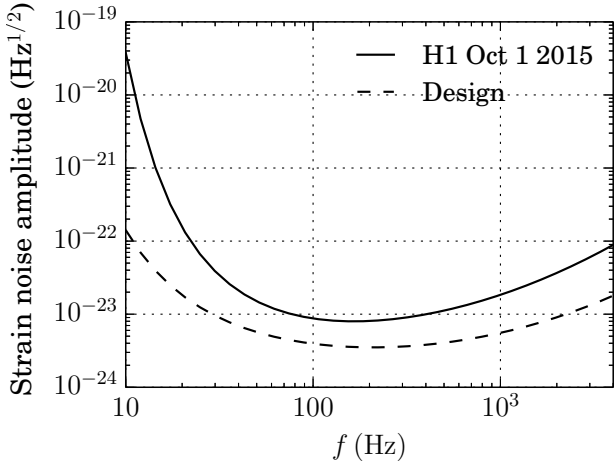}
 \end{center}
\caption{Sensitivity spectrum obtained from LIGO H1 on Oct 1 2015 (LIGO-G1501223) \citep{Kissel15} and the design sensitivity spectrum \citep{2016LRR....19....1A}.\label{fig:sensitivity}}
\end{figure}

Instead of using $S_{\rm N}(f)$ and $f_{\rm off}$ adopted in 
\citet{2013MNRAS.435.1358T}, we adopt the following functions.
For $S_{\rm N}(f)$, we perform
a least square fitting to the latest sensitivity spectrum of
aLIGO-Hanford on October 1, 2015 \citep{Kissel15}. The fitted 
sensitivity spectrum is given by
\begin{eqnarray}
\log ({\rm Strain}) &=& -24 + 4.4[\log(f)]^{-3.0} + 0.034[\log (f)]^{3.1}
\end{eqnarray}
and shown in Figure \ref{fig:sensitivity}.
We note that $S_{N}(f)=({\rm Strain})^2$.
We also perform a fitting to the final design spectrum 
\citep{2016LRR....19....1A} and obtain
\begin{eqnarray}
 \log ({\rm Strain}) &=& -24 + 2.2[\log(f)]^{-2.3} + 0.011[\log (f)]^{3.7}.
\end{eqnarray}
This function is also shown in Figure \ref{fig:sensitivity}. 
We adopt this only for $D_{\rm L,0}=200$ Mpc. 
For $f_{\rm off}$, we adopt the ringdown frequency expressed as:
\begin{eqnarray}
f_{\rm off} = \frac{c^3}{3\pi G M_{\rm f}}[1.5251-1.1568(1-a_{\rm f})^{0.1292}],
\end{eqnarray}
where $M_{\rm f}$ and $a_{\rm f}$ is the final mass and spin of the merger
remnant \citep{2006PhRvD..73f4030B,2016MNRAS.457.4499H}. We assume 
$a_{\rm f}=1$ and $M_{\rm f}=M_{\rm f, max}=0.89(m_{1}+m_{2})$ 
\citep{2014PhRvD..90j4004H}. 

For $M_{\rm ch,0}$ and $f_{\rm off, 0}$,
we adopt the chirp mass and cut-off frequency for an
NS-NS merger with a mass of $1.4M_{\odot}$, 
which are $1.2M_{\odot}$ and 4800\,Hz, respectively.
Here, $D_{\rm L,0}$ is the detection limit in luminosity distance
for a NS-NS merger. We adopt $D_{\rm L,0}=$40, 80, 120, and 
200 Mpc, which correspond to the detectable range for the current
and future observations. The detectable range for NS-NS mergers
is expected to be 
40--80\,Mpc for the observation in 2015--16, 80--120\,Mpc in 2016--2017, 
120--170\,Mpc in 2017--18, and 200\,Mpc for the full sensitivity
in 2019. The signal-to-noise ratio (SNR) for detection is assumed to be 
8 \citep{2016LRR....19....1A}. Using our description, the 
cut-off frequency and the SNR of GW150914 are calculated 
as 190 Hz and 21, respectively, assuming $D_{\rm L,0}=80$Mpc, 
$z=0.09$, and the black hole masses are 36 and $29M_{\odot}$ 
\citep{2016PhRvL.116f1102A}. These values are consistent with the 
actual observation.

\begin{figure}
 \begin{center}
  \includegraphics[clip, width=1.\columnwidth]{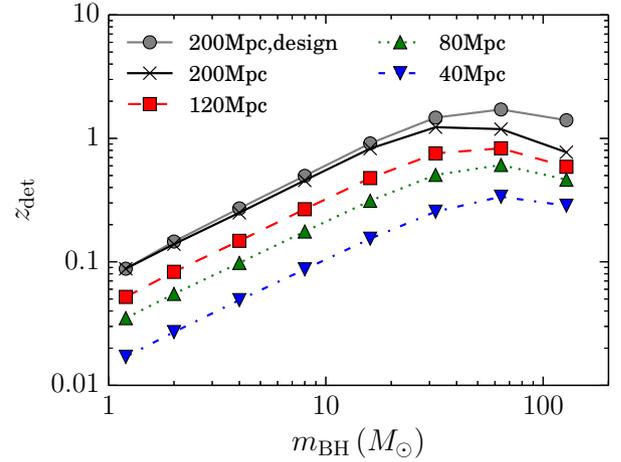}
 \end{center}
\caption{The detection limit as a function of the mass of a BH for $D_{\rm L,0}=200$, 120, 80, and 40 and for each sensitivity spectrum. Here we assume that BBHs are equal mass systems. The smallest value of $m_{\rm BH}$ corresponds to $1.2M_{\odot}$. \label{fig:detectable_z}}
\end{figure}

In Figure \ref{fig:detectable_z}, we present the maximum detectable 
redshift ($z_{\rm det}$) as a function of the mass of NSs or BHs 
($m_{\rm BH}$) for each $D_{\rm L, 0}$ and sensitivity spectrum. Here 
we assume that the BBHs are equal mass systems and that SNR$>8$ for 
the detection. These results are 
roughly consistent with obtained in other works 
\citep[e.g., see Figure 4 in][]{2016ApJ...818L..22A}.

We calculated the cumulative detection rate up to $z$ as 
\begin{eqnarray}
\Gamma _{\rm det}(<z) = f_{\rm det}^{-3}\int _{0}^{z} \left[ n_{\rm GC}(z')
\frac{dV(z')}{dz'} \frac{\Gamma_{\rm det}(z')}{1+z'}\right] dz', 
\end{eqnarray}
where $f_{\rm det}=2.26$ is a factor for the non-uniform pattern of 
detector sensitivity and random sky orientation of sources
\citep{1993PhRvD..47.2198F}, and $\Gamma_{\rm det}(z')$ is 
the detection rate at $z'$, which is the rate of mergers satisfying 
$C_{\rm det}>1$.

We estimate the detection rates of BBH mergers by aLIGO, 
using models for the distribution of BBHs based on direct $N$-body 
simulations \citep{2013MNRAS.435.1358T}, cosmic star-cluster formation 
history, and the detection model described in this section. We first 
generate BBH merger history per cluster using the models of 
\citet{2013MNRAS.435.1358T}.
Following the cosmic star-cluster formation rate, we assume that 
star clusters form every 1\,Gyr. From these, we calculate 
the merger rate of BBHs in the universe and their mass distribution. 
Assuming the detection model as is described in this section, we 
further calculate the detection rates of BBHs and their mass 
distribution.
We performed these calculations using Python scripts working on AMUSE 
(the Astrophysical Multipurpose Software Environment) framework
\citep{2009NewA...14..369P,2013A&A...557A..84P}.

\section{Results: Detection rates}

\begin{figure}
 \begin{center}
  \includegraphics[clip, width=1.\columnwidth]{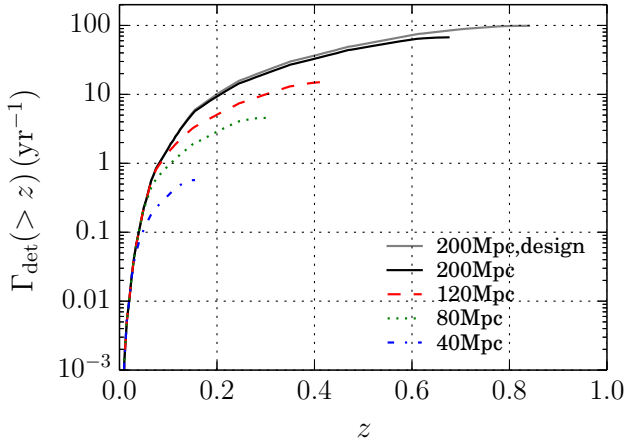}
 \end{center}
\caption{Cumulative detection rate per year within redshift $z$ for $D_{\rm L,0}=200$, 120, 80, and 40 and for each sensitivity spectrum.\label{fig:det_rate}}
\end{figure}

In Figure \ref{fig:det_rate}, we present the cumulative
detection rate of BBH mergers as a function of redshift ($z$)
obtained from our models. 
The total detection rates are 67, 15, 4.6,
and 0.57 per year 
for $D_{\rm L,0}$=200, 120, 80, and 40\,Mpc, respectively, assuming 
the sensitivity spectrum on Oct 1 in 2015 and $T=10$.
Even for the current detection limit
($\sim 70$\,Mpc) \citep{Kissel15}, the detection rate reaches 
several per year. Here, we assume that massive star clusters continue
to form to $z=0$ following the cosmic star cluster formation 
history. 
With the final design sensitivity spectrum, 
the total detection rate increases to 99 per year
for $D_{\rm L,0}=200$\,Mpc. If we assume a larger value for
$T$, the detection rate increases.

\begin{figure}
 \begin{center}
  \includegraphics[clip, width=1.\columnwidth]{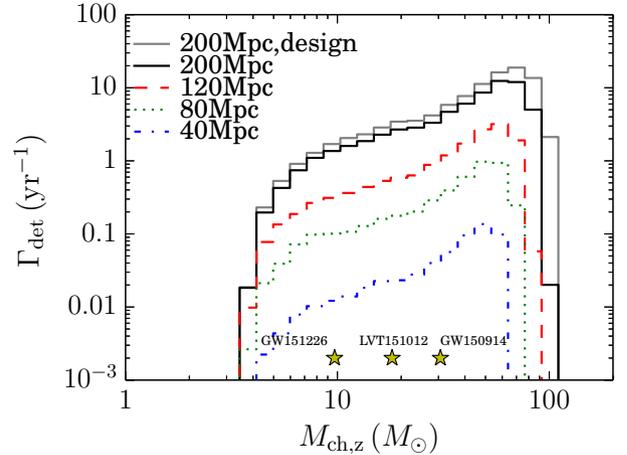}
 \end{center}
\caption{Redshifted chirp mass distribution of detected merging BBHs for the detection range ($D_{\rm L, 0}$) of 200, 120, 80, and 40\,Mpc (black full, red dashed, green dotted, and blue dash-dotted lines), respectively. Gray full line indicate $D_{\rm L, 0}=200$ Mpc, but for the final design sensitivity spectrum (see Figure \ref{fig:sensitivity}) Note that $D_{\rm L,0}=80$\,Mpc corresponds to the sensitivity of current aLIGO. Stars indicate the mass of detected BBH mergers \citep{2016arXiv160604856T}. \label{fig:det_MF}}
\end{figure}

We present the redshifted chirp mass function of the detected
BBH mergers in Figure \ref{fig:det_MF}.
The mass function has a peak at $M_{\rm ch, z}\sim $60--80\,$M_{\odot}$
slightly depending on the values of $D_{\rm L,0}$.
The merger rates of the intermediate mass range of the BBHs 
($M_{\rm ch, z}\sim$10--30) is almost flat. We also plot
the mass of the detected BBH mergers with redshifted chirp masses 
of 31, 9.7, and 18 for GW150914, GW151226,
and LVT151012, respectively \citep{2016arXiv160604856T}. They are 
located in this flat region.

\begin{figure*}
 \begin{center}
  \includegraphics[clip, width=0.9\columnwidth]{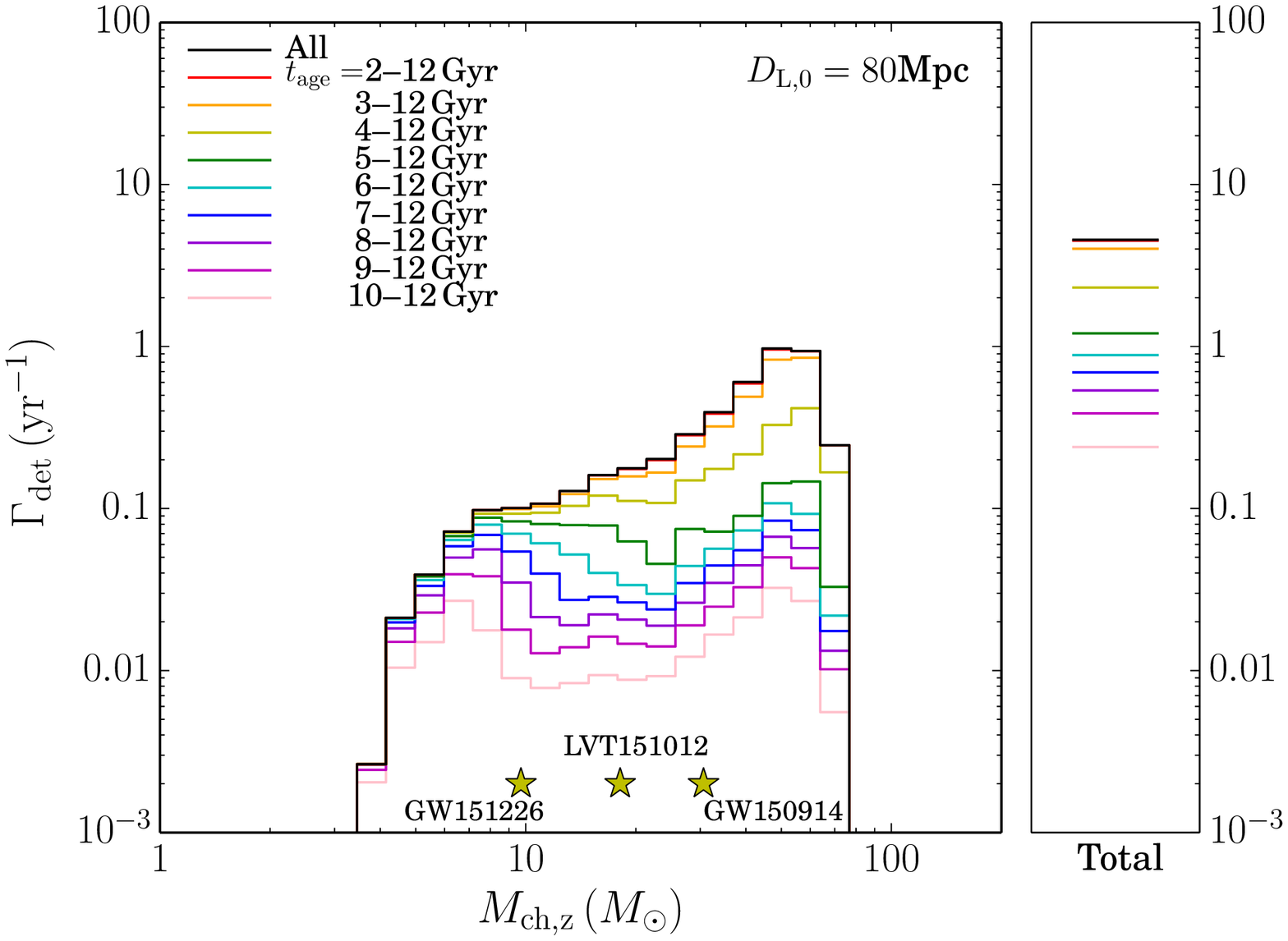}
  \includegraphics[clip, width=0.9\columnwidth]{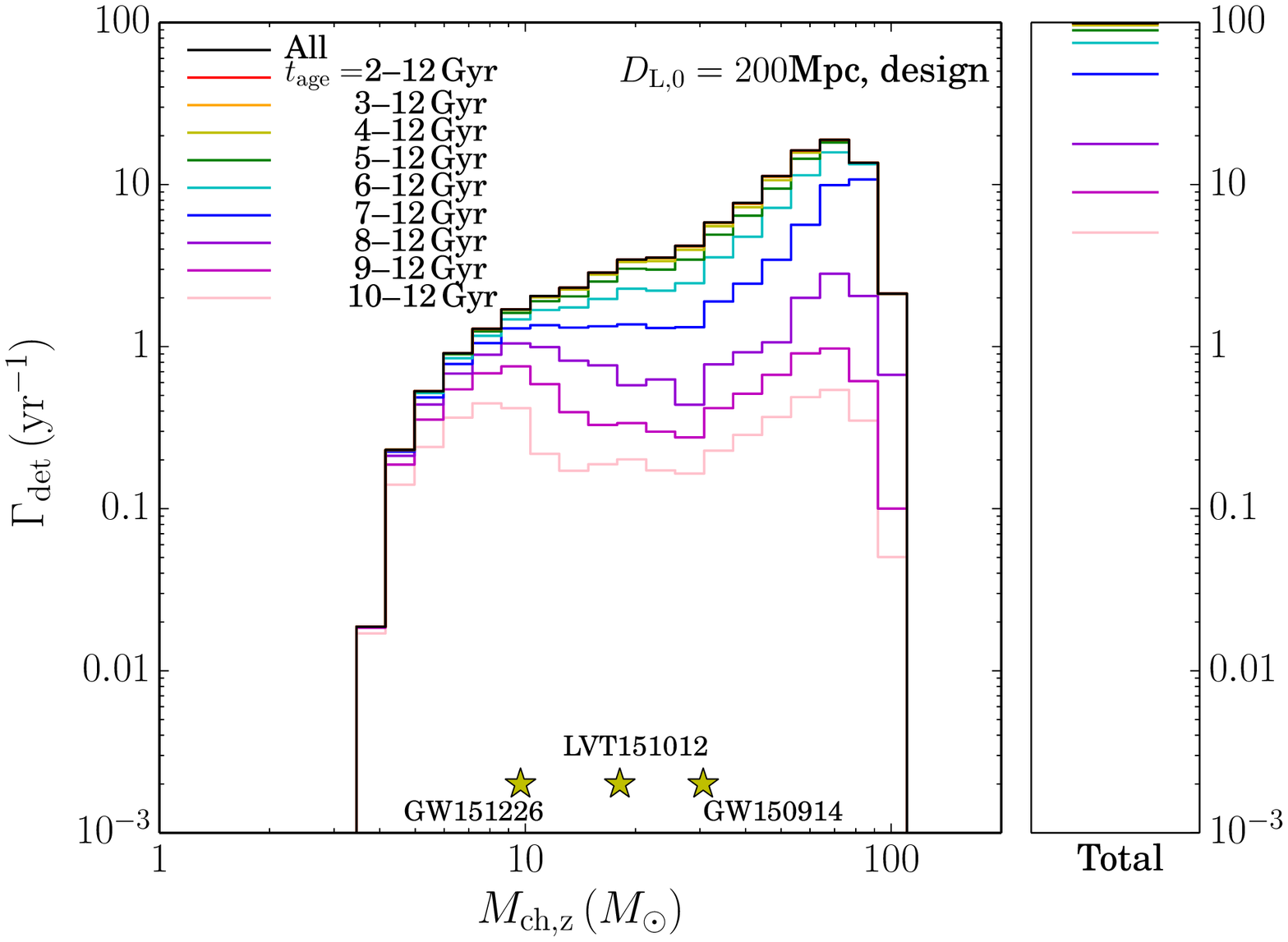}
 \end{center}
\caption{Redshifted chirp mass distribution of detected merging BBHs for the ranges of cluster formation for the current detection limit of aLIGO (left) and future detection limit (right).\label{fig:detection_mass_func}}
\end{figure*}

Figure \ref{fig:detection_mass_func} shows the changes in detection rates 
when we limit the formation epoch of star clusters. 
The left and right panels are for the cases of current aLIGO detection
limit ($D_{\rm L,0}=80$\,Mpc) and future detection limit
($D_{\rm L, 0}=200$ Mpc), respectively.
Black curves correspond to the detection
rates shown in Figure \ref{fig:det_MF}.
If we consider star clusters older than 10 Gyr, the detection
rates decrease down to 0.23 and 5.1 per year for the current 
and future detection limits, respectively. Not only the 
detection rates, the shapes of the 
redshifted chirp-mass function also changes. 
The detection rates of higher-mass BBH mergers significantly
decreases and the shape of the mass function becomes double-peaked,
because massive BBH mergers originating from younger star 
clusters are excluded. The low-mass peak is located at 7--8$M_{\odot}$.
We summarize the merger rate density and detection 
rates for the current and future detection limits of LIGO in 
Table \ref{tb:results}.

\begin{figure}
 \begin{center}
  \includegraphics[clip, width=1.\columnwidth]{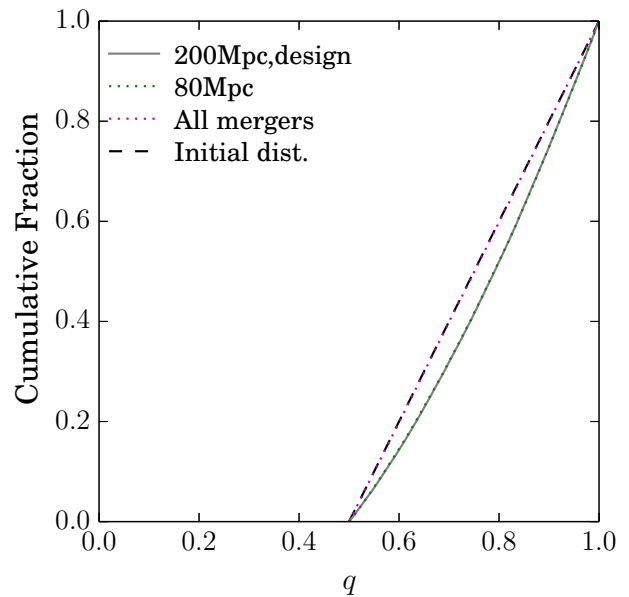}
 \end{center}
\caption{Cumulative distribution of BH mass ratio of detected BBH 
mergers for models with $D_{\rm L,0}=200$\,Mpc and $D_{\rm L, 0}=80$\,Mpc, all merger events, and the initial distribution (see equation (\ref{eq:q})) \label{fig:q}}
\end{figure}

In Figure \ref{fig:q}, we also present the distribution of BH mass 
ratio ($q$) of detected BBH mergers. The distribution of $q$ does
not depend on the detectable distances. We therefore plot it only
for models with $D_{\rm L,0}=200$\,Mpc and the designed sensitivity 
and current aLIGO ($D_{\rm L,0}=80$\,Mpc). We also plot the distribution
of $q$ for all merger events occurred until the current age of the 
universe and confirmed that this distribution of $q$ is caused by
the dependence of $t_{\rm GW}$ on $q$. BBHs with a higher eccentricity
have a shorter $t_{\rm GW}$, and therefore the $q$-distribution of 
merging BBHs are more dominated by eccentric binaries compared to 
the initial distribution.
The distribution of the mass ratio predicted by common envelope scenario,
which depends on the initial distribution of binary mass ratio and 
the metallicity, has a much steeper dependence on the mass ratio 
\citep{2014ApJ...789..120B,2015ApJ...806..263D},
although this may be because they assumed a flat initial mass ratio 
($q\sim 1$ for most of massive stars) distribution.
Even if primordial binaries in star clusters have the same initial
mass ratio, the mass ratio would change due to the dynamical evolution
in star clusters.
Thus, we can probably regard the shapes of BBH mass function and the 
mass ratio as the signature of the
dynamical formation of BBHs in star clusters and expect that 
observations of a large number of BBH mergers in the near future will 
tell us the formation mechanisms of merging BBHs.

\begin{table*}
  \tbl{Summary of the results}{%
  \begin{tabular}{llccc}
\hline
   Model &  & Merger rate density ($z<0.1$)  &  \multicolumn{2}{c}{Detection rate}\\
         &  & (Gpc$^{-3}$\,yr$^{-1}$)  & Current (yr$^{-1}$) & Future (yr$^{-1}$)\\\hline
   $M_{\rm BH, max}=54M_{\odot}$ & $t_{\rm age}=$all & 57 & 4.6 & 99 \\
                             & $t_{\rm age}=$10--12\,Gyr & 13 & 0.23 & 5.1 \\
                             & $t_{\rm age}=$all, with natal kick & 16 & 3.9 & 86 \\
                             & $t_{\rm age}=$10--12\,Gyr, with natal kick & 1.3 & 0.14 & 2.9\\
   $M_{\rm BH, max}=20M_{\odot}$ & $t_{\rm age}=$all & 57 & 1.8 & 37 \\
                             & $t_{\rm age}=$10--12\,Gyr & 14 & 0.17 & 3.7 \\ 
\hline
  \end{tabular}}\label{tb:results}
  \begin{tabnote}
    $M_{\rm BH, max}$ is the maximum mass of BHs. $t_{\rm age}$ indicates the range of cluster age we included. Current and future detection rates indicate the detection rate for $D_{\rm L,0}=80$ and 200\,Mpc, respectively. 
  \end{tabnote}
\end{table*}

\section{Discussion}

\subsection{Natal kicks}

BHs receive natal kicks when they form because of asymmetric supernova
explosions. The kick velocities can exceed escape velocities of 
star clusters, and as a consequence the kicked BHs can be ejected
from the host star clusters. 
The observation of radio pulsars suggests the one-dimensional
velocity dispersion of 265 km\,s$^{-1}$ \citep{2005MNRAS.360..974H},
which is much larger than typical escape velocities.
On the other hand, recent stellar evolution models suggest that 
the kick velocity depends on the fraction of fallback materials
when they explode and that
massive BHs formed via direct collapse do not receive any kicks
\citep{2006ApJ...650..303B,2012ApJ...749...91F}.

\citet{2013MNRAS.435.1358T} performed direct $N$-body simulations 
assuming a constant retention fraction between 0.25 and 1.0 
irrespective of BH masses and found that the BBH merger rates are 
proportional to the retention fraction. 
In this section, we estimate the merger and detection rates of 
BBHs assuming that BBH merger rate is proportional to 
the retention fraction and that the retention fraction depends
on BH masses.
For simplicity, we assume that the retention fraction linearly
changes from 0.1 for $M_{\rm BH}=3M_{\odot}$ to 1.0 for
$M_{\rm BH}=20M_{\odot}$ and that all BHs with $M_{\rm BH}>20M_{\odot}$
remain in the clusters. 
The minimum BH mass for the direct collapse depends on metallicity
and $M_{\rm BH}>$15--30$M_{\odot}$ for $Z=0.01Z_{\odot}$, where $Z_{\odot}$
is the solar metallicity, is suggested \citep{2016Natur.534..512B}.

\begin{figure*}
 \begin{center}
  \includegraphics[clip, width=0.9\columnwidth]{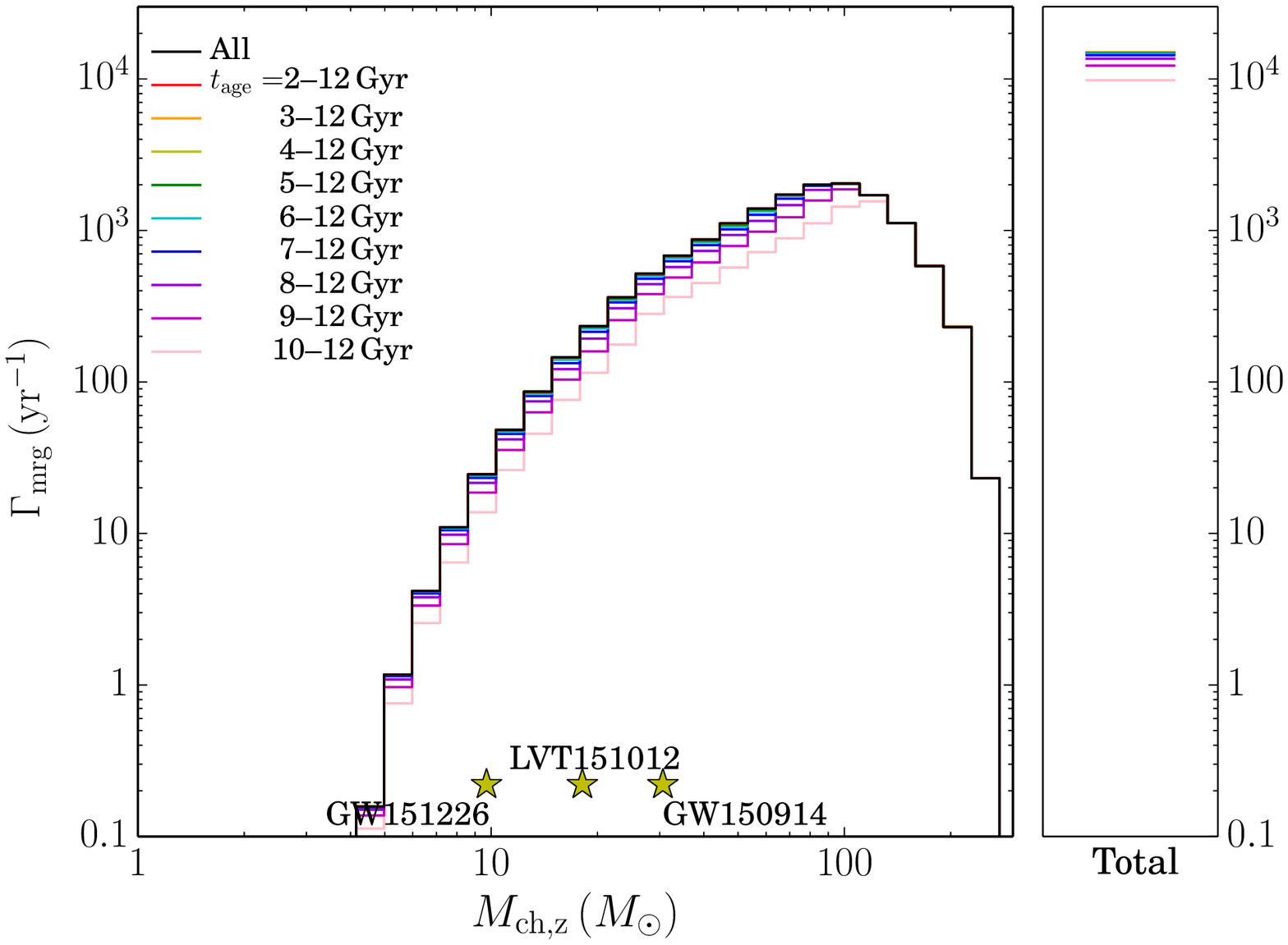}
  \includegraphics[clip, width=0.9\columnwidth]{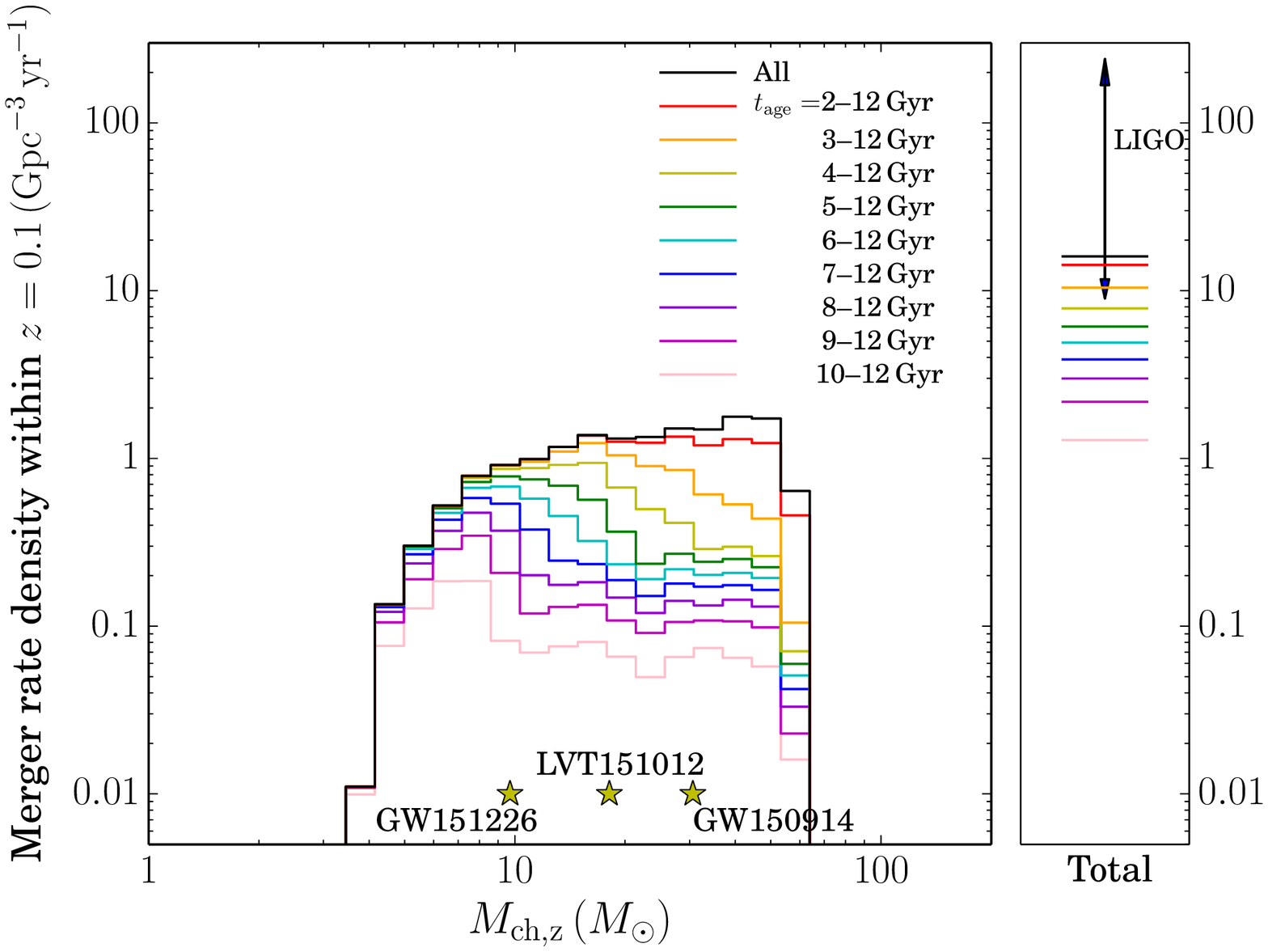}
 \end{center}
\caption{Same as Figure \ref{fig:merger_mass_func}, but with the natal-kick 
model. \label{fig:merger_mass_func_kick}}
\end{figure*}

In Figure \ref{fig:merger_mass_func_kick}, we present the merger rate
of all mergers and merger rate density for the local universe ($z<0.1$)
as a function of 
redshifted chirp mass. Compared with the case without kicks (see 
Figure \ref{fig:merger_mass_func}), the fraction of low-mass BHs 
significantly decreases, and the total merger rates also decrease. 
The total retention fraction of BHs obtained from our assumption 
is $\sim0.7$, which is consistent with retention fractions of 
0.4--0.7 obtained from models of kick velocities depending on 
BH masses (the amount of fallback materials) 
\citep{2006ApJ...650..303B,2007MNRAS.374...95P}.
The merger rate density within $z=0.1$ including young star clusters 
born down to $z=0$ is 16\,Gpc$^{-3}$\,yr$^{-1}$, which is still 
within the range estimated by LIGO \citep{2016arXiv160604856T}.
If young star clusters are excluded, the merger rate density 
drops to 1.3\,Gpc$^{-3}$\,yr$^{-1}$.
The results with natal kicks are also summarized in 
Table \ref{tb:results}.

\begin{figure*}
 \begin{center}
  \includegraphics[clip, width=0.9\columnwidth]{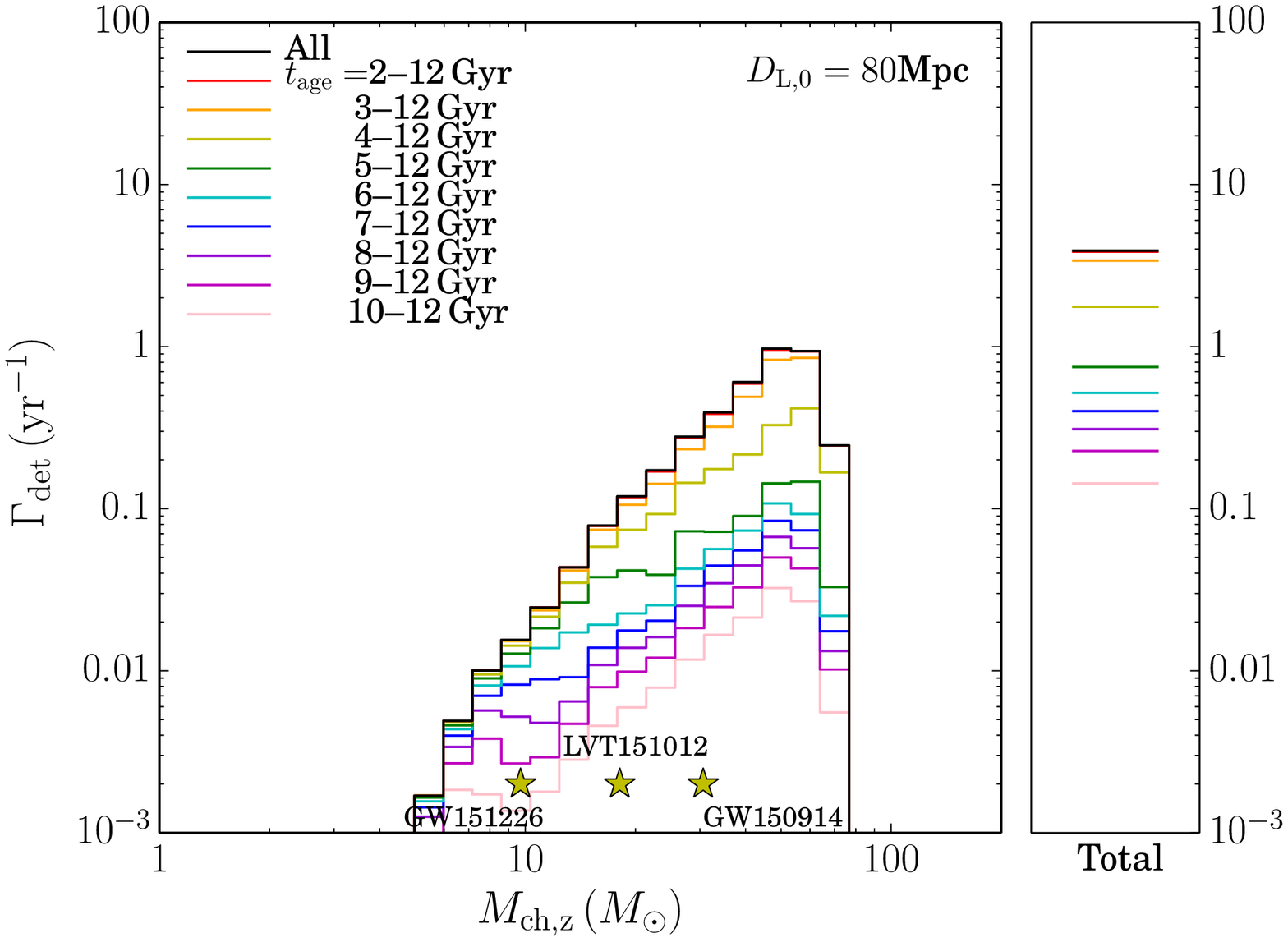}
  \includegraphics[clip, width=0.9\columnwidth]{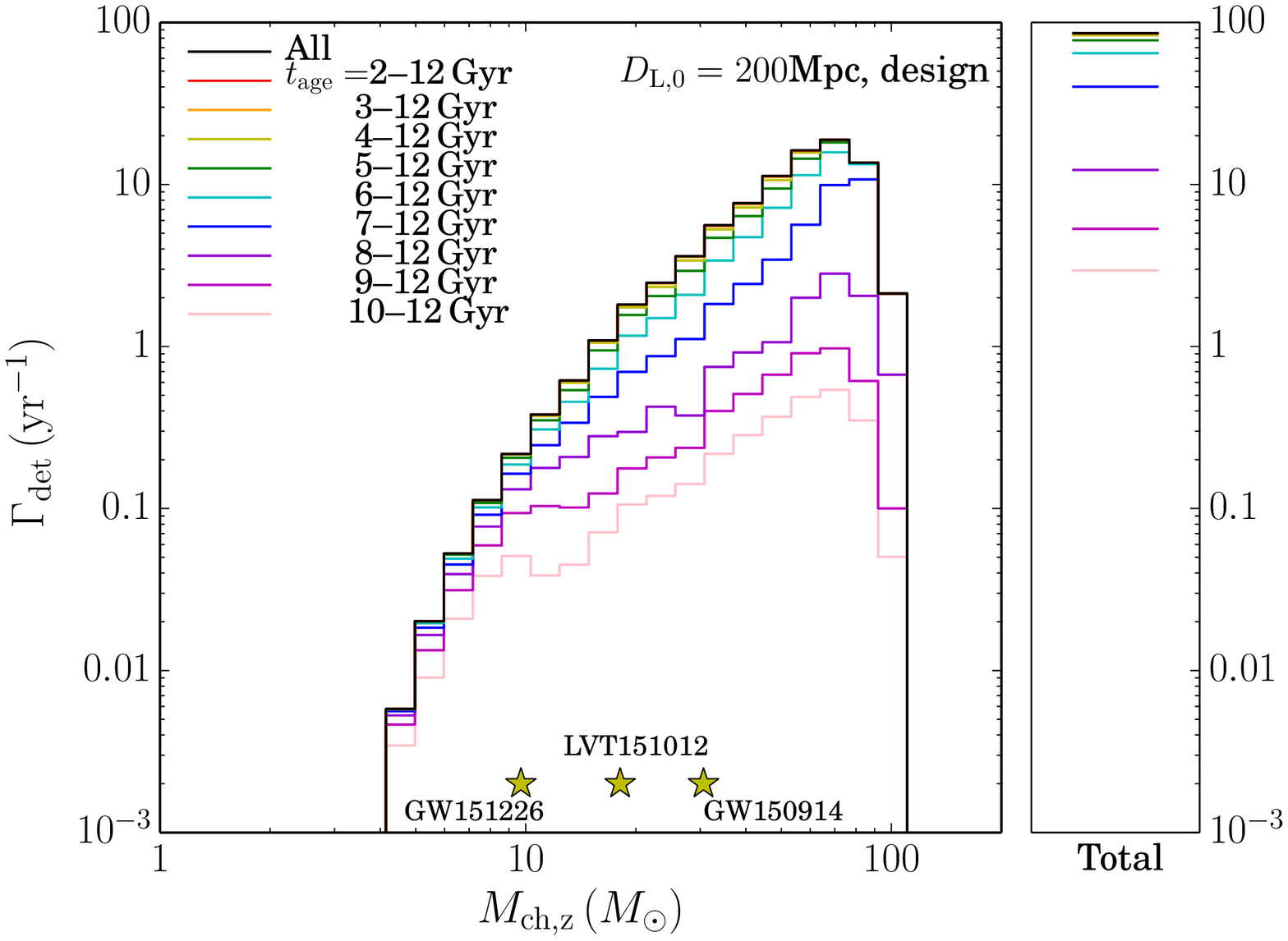}
 \end{center}
\caption{Same as Figure \ref{fig:detection_mass_func}, but with the natal-kick 
model. \label{fig:detection_mass_func_kick}}
\end{figure*}

Figure \ref{fig:detection_mass_func_kick} shows the distribution of 
detection rates as a function of redshifted chirp mass
expected from our natal-kick model. 
The estimated detection rates are 
0.14--3.9 per year for the current detection limit ($D_{\rm L, 0}=80$Mpc)
2.9--86 per year for the future detection limit ($D_{\rm L, 0}=200$Mpc).
In the case without natal kicks, the 
distribution of detection rates show a peak close to the high-mass 
end, and another peak appears when younger clusters are ignored 
(see Figure \ref{fig:detection_mass_func}). With natal kicks, however,
the detection rate always show a strong peak at the high-mass end.

\subsection{Metallicity evolution and BH mass function}

Another important parameter we had not accounted for is
the change of the metallicity of star clusters as a function of $z$ and 
the resulting BH mass function.   
The metallicity strongly affect the stellar evolution and the 
final BH masses. Stars with lower metallicity can form 
more massive BHs. Recent stellar evolution models suggest
that the maximum BH mass reaches $\sim 100M_{\odot}$ for $Z=0.0001$,
but only $\sim 15M_{\odot}$ for the solar metallicity
\citep{2016Natur.534..512B}. The mean metallicity of the universe 
increases with time \citep{2014ARA&A..52..415M}. The 
maximum mass of BHs, therefore, should decrease as the formation epoch of 
the host star clusters delays. Furthermore, the metallicity of 
star clusters born in the same redshift should have a 
dispersion. In the local universe, starburst galaxies which are 
still forming massive star clusters, such as 
the Large and Small Magellanic Clouds, have a metallicity 
lower than the average ($\sim 0.4$ and $\sim 0.2$, respectively) 
\citep{2005AJ....129.1465C,1998AJ....115..605L}. Recent stellar
evolution models expected that BHs with a maximum mass of 
$\sim 30M_{\odot}$ \citep{2016Natur.534..512B} may form in the
Magellanic Clouds.  
In addition, the natal kick velocity also decreases as the 
metallicity decreases \citep{2016Natur.534..512B}. Thus,
evolution models of the mean metallicity with a dispersion should 
be included, but has not fully been done yet for star cluster models.
\citet{2016arXiv160202444R} recently performed a series of Monte-Carlo
simulations of star clusters for three different metallicity 
(0.01, 0.05, and 0.025\,$Z_{\odot}$) with recent prescriptions of stellar 
evolution and natal kicks. They estimated BBH merger rates, but they 
assumed all star clusters were born 12\,Gyr ago.

Since the metallicity affects the stellar evolution and kick velocity,
the dynamical evolution of star clusters may be 
affected. In some cases, however, the effects  on the dynamical 
evolution is suggested to be limited due to 
the small fraction of massive stars \citep{2013MNRAS.435.1358T},
although we still need to investigate  the effect by performing direct $N$-body
simulations changing the metallicity and related stellar 
evolution models. Such simulations consume a large amount of 
computational resources. We therefore do not perform additional 
direct $N$-body simulations here, but test a model with a smaller 
upper-mass limit of BH mass based on the model of 
\citet{2013MNRAS.435.1358T}.

\begin{figure*}
 \begin{center}
  \includegraphics[clip, width=0.9\columnwidth]{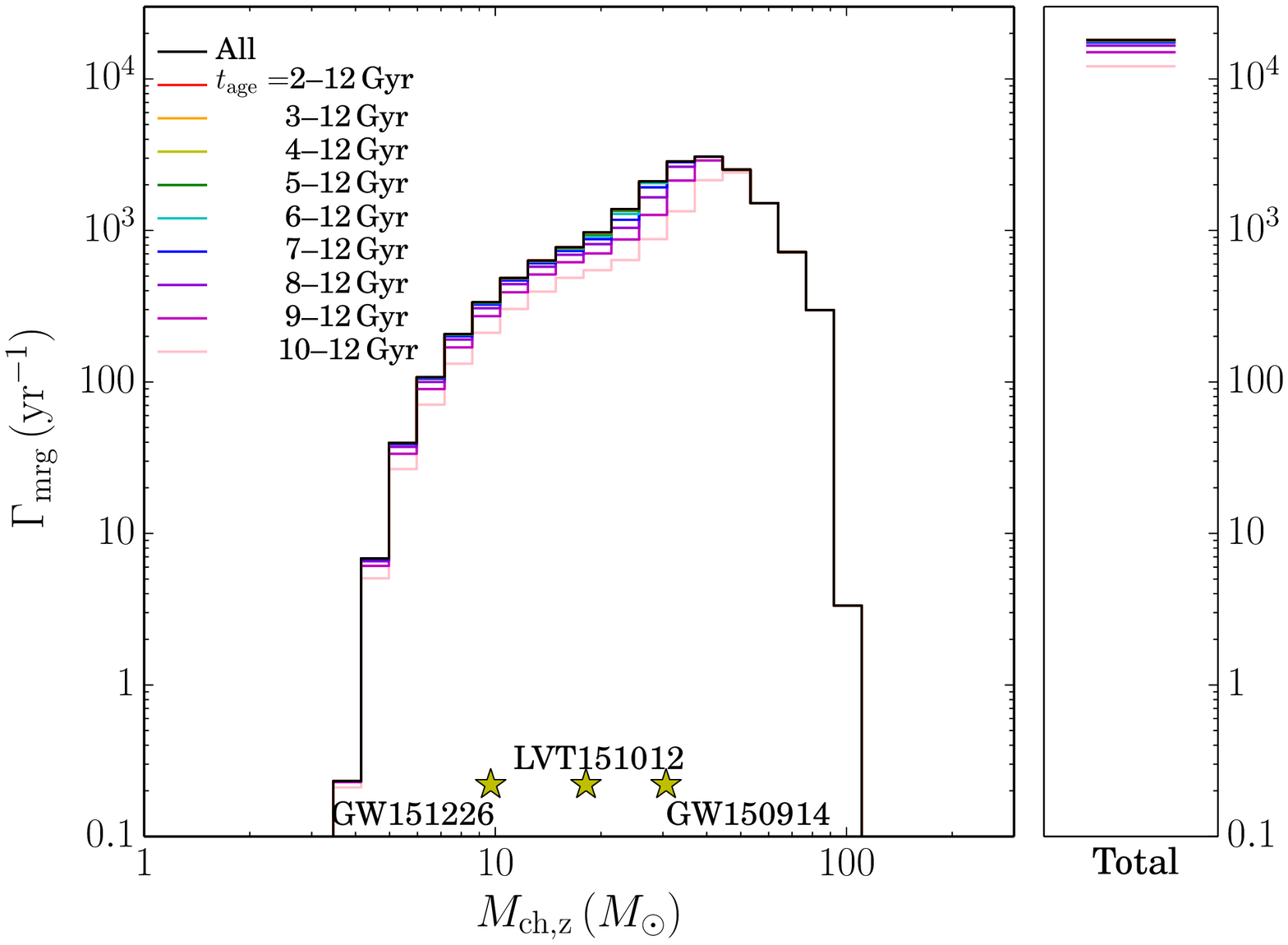}
  \includegraphics[clip, width=0.9\columnwidth]{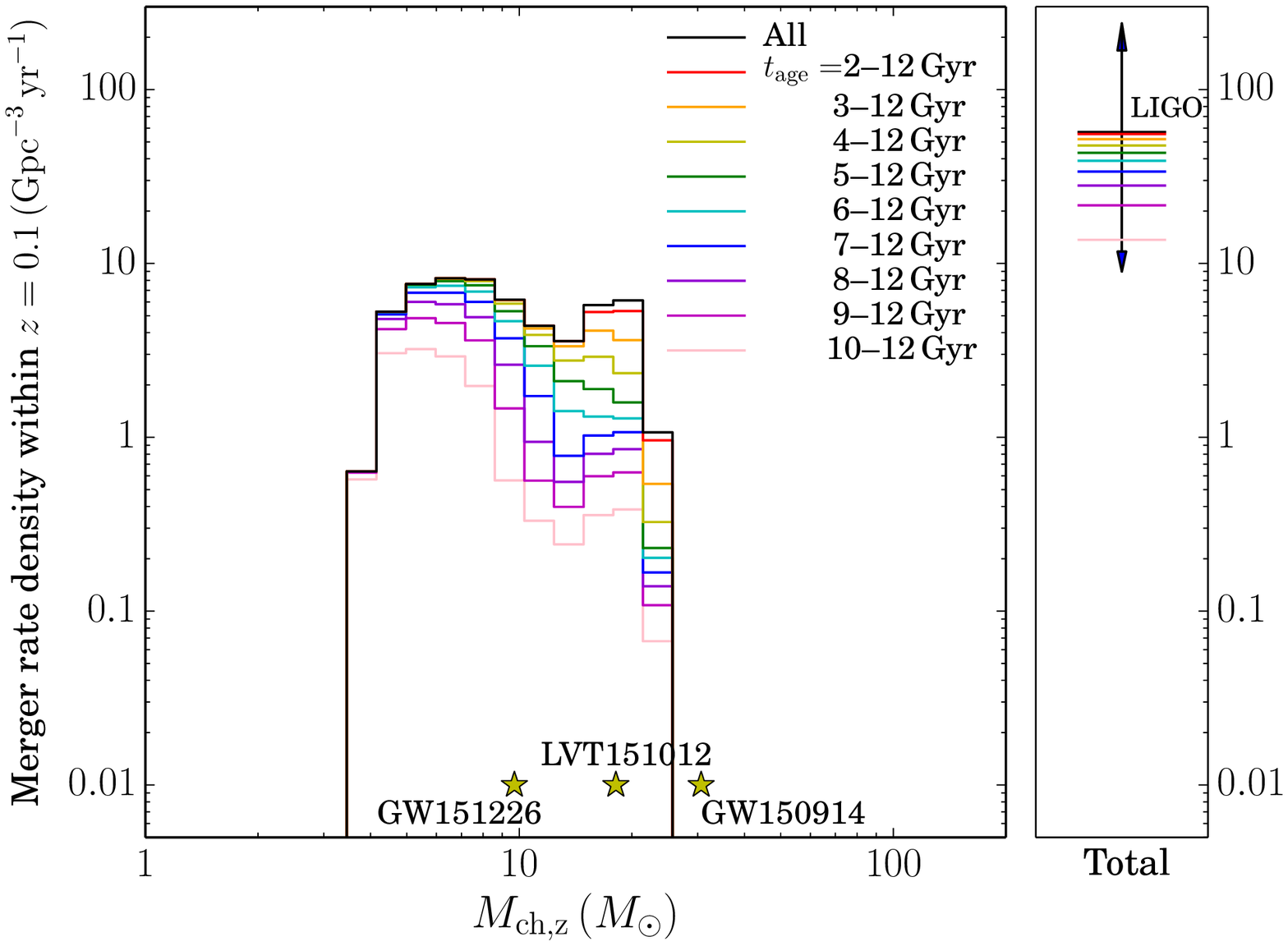}
 \end{center}
\caption{Same as Figure \ref{fig:merger_mass_func}, but for a model with a maximum BH mass of $20M_{\odot}$. \label{fig:merger_mass_func_M20}}
\end{figure*}

Instead of considering a cosmic metallicity evolution and the 
dispersion, we perform the same analyses for a model with 
a maximum BH mass of $20M_{\odot}$, which is obtained from
the direct $N$-body simulations of
\citet{2013MNRAS.435.1358T} and compare the results to the cases
with a maximum BH mass of $54M_{\odot}$. 
In Figure \ref{fig:merger_mass_func_M20}, we present the merger rate
of all mergers and merger rate density for $z<0.1$ as a function of 
redshifted chirp mass. Since there is no massive BHs than $20M_{\odot}$,
no GW150914-like BBH merger occurs. The shape of mass functions 
are similar to those for a model with $m_{\rm BH, max}=54M_{\odot}$,
and the merger rate densities within $z=0.1$ are 14--57\,Gpc$^{-3}$\,yr$^{-1}$
changing the ranges of cluster formation epoch.

\begin{figure}
 \begin{center}
  \includegraphics[clip, width=1.\columnwidth]{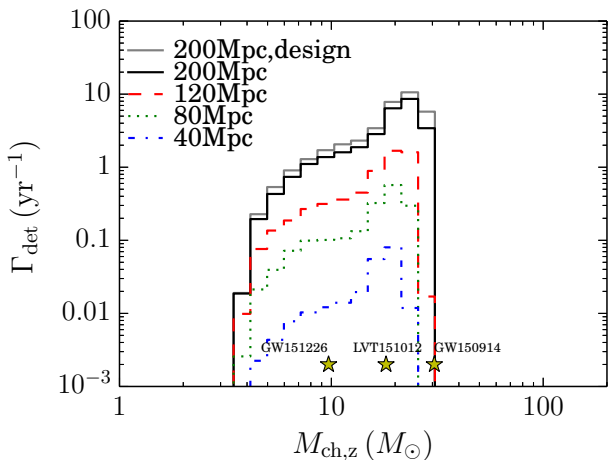}
 \end{center}
\caption{Same as Figure \ref{fig:det_MF}, but for $m_{\rm BH, max}=20M_{\odot}$.
\label{fig:det_MF_M20}}
\end{figure}

In Figure \ref{fig:det_MF_M20}, we present the distribution of 
detection rate as a function of redshifted chirp mass.
The detection rates are 0.22, 1.8, 6.0, 29, 
and 37\,yr$^{-1}$ for $D_{\rm L,0}=$40, 80, 120, 200\,Mpc and 200\,Mpc
with designed sensitivity, respectively. These results are 
summarized in Table \ref{tb:results}.
Compared with the results of \citet{2013MNRAS.435.1358T},
our model enhances massive BBH mergers. While \citet{2013MNRAS.435.1358T}
assumed star cluster formation at 10 or 12\,Gyr ago, we 
adopt a star-cluster formation history including
younger star clusters. Since more massive BBHs merge more frequently
in younger ages of star clusters, most of massive BBH mergers occur
at high $z$ and thus are unreachable with current or future LIGO, if we assume single cluster formation event 
at $>10$\,Gyr ago. Our redshifted chirp mass function, therefore,
contains massive BBH mergers more than that of \citet{2013MNRAS.435.1358T}.   
We also updated the detection criterion using 
the recent model of the LIGO sensitivity. This also changed the 
shape of mass function of detected BBHs from that of 
\citet{2013MNRAS.435.1358T}.

\section{Summary}

We estimated the BBH merger rates and the detection rates originating from
the dynamical evolution of star clusters using
a model of BBH-merger distribution obtained from $N$-body simulations performed 
by \citet{2013MNRAS.435.1358T}.
We assumed that globular clusters were born following the cosmic
star formation rate. Combining the BBH merger history per cluster and the 
cosmic star-cluster formation history, we obtained the merger 
rate density in the local universe ($z<0.1$) of 57\,Gpc$^{-3}$\,yr$^{-1}$
including young star clusters and 13\,Gpc$^{-3}$\,yr$^{-1}$ taking 
old globular clusters born 10--12 Gyr ago into account. These values 
are consistent with the values estimated from LIGO  
\citep[9--240 Gpc$^{-3}$\,yr$^{-1}$,][]{2016arXiv160604856T}. 
We also estimated the 
detection rates of BBH merger events as 67, 15, 4.6, 
and 0.67 per year 
for $D_{\rm L,0}$=200, 120, 80, and 40\,Mpc, respectively, assuming 
the sensitivity spectrum on Oct 1 in 2015 and including young star 
clusters. 
For the final design sensitivity spectrum, 
the total detection rate increases to 99 per year
for $D_{\rm L,0}=200$\,Mpc. If we assume that only star clusters 
born in a higher redshift can form BBHs, this value drops to 
$\sim 5$ per year. In addition, if we assume that less massive 
BHs are ejected due to the natal kicks, it decreases down to 
$\sim 3$ per year. Taking natal kicks into account, 
the merger rate density in the local universe is estimated to be 
1.3 and 16\,Gpc$^{-3}$\,yr$^{-1}$ excluding and including young 
star clusters, respectively.

We also predicted the redshifted 
chirp mass distribution of detected BBH mergers.
The mass function of detected BBH mergers dynamically formed in
star clusters has a peak at its high-mass end, if we account for 
all star clusters born down to $z=0$. The detection rate in 
the intermediate mass range is almost flat. If we assume that
only star 
clusters born in a higher redshift can form BBHs, the mass
distribution changes to a double-peaked shape with a low-mass
peak at 7--8$M_{\odot}$.
The high-mass peak is not because of the initial mass
function of BHs, but because of the dynamical evolution of star 
clusters and the typical lifetime of BBHs
after their formation in star clusters. 
In dense star clusters, massive BHs selectively form binaries,
and their separations shrink via three-body encounters. 
Thus, a mass distribution with a peak at its high-mass end 
is a typical signature of the dynamically formed BBHs and 
is also predicted by 
similar studies \citep[e.g.,][]{2016arXiv160202444R}.

\begin{ack}
We thank the anonymous referee for the useful comments.
This work was supported by JSPS KAKENHI Grant Number 26800108 and 
17H06360. 

\end{ack}

\bibliography{reference}

\end{document}